\documentclass[conference]{IEEEtran}

\usepackage[utf8]{inputenc}
\usepackage[T1]{fontenc}
\usepackage[english]{babel}
\usepackage{xspace}
\usepackage{cite}
\usepackage{url}
\usepackage{amsmath}
\usepackage{amssymb}
\usepackage{amsthm}
\usepackage{algorithm}
\usepackage{flushend}
\ifCLASSINFOpdf
  \usepackage[pdftex]{graphicx}
\else
  \usepackage[dvips]{graphicx}
\fi

\usepackage{color}
\usepackage[noend]{algpseudocode}
\usepackage[caption=false,font=footnotesize,margin=0pt,farskip=0pt]{subfig}

\newtheorem{definition}{Definition}

\algnewcommand\algorithmicdata{\textbf{Data:}}
\algnewcommand\Data{\item[\algorithmicdata]}

\newcommand{\accio}{\textsc{ALP}\xspace}

\newcommand{\promesse}{\textsc{P\footnotesize{ROMESSE}}\xspace}
\newcommand{\geoi}{\textsc{G\footnotesize{EO}\normalsize{-I}}\xspace}

\begin{document}

\title{Adaptive Location Privacy with \textsc{ALP}}

\author{\IEEEauthorblockN{Vincent Primault, Antoine Boutet, Sonia Ben Mokhtar, Lionel Brunie}
\IEEEauthorblockA{Université de Lyon, CNRS\\
INSA-Lyon, LIRIS, UMR5205, F-69621, France\\
\{vincent.primault,antoine.boutet,sonia.ben-mokhtar,lionel.brunie\}@liris.cnrs.fr}}

\maketitle

\begin{abstract}
With the increasing amount of mobility data being collected on a daily basis by location-based services (LBSs) comes a new range of threats for users, related to the over-sharing of their location information.
To deal with this issue, several location privacy protection mechanisms (LPPMs) have been proposed 
in the past years.
However, each of these mechanisms comes with different configuration parameters that have a direct impact both on the privacy guarantees offered to the users and on the resulting utility of the protected data.
In this context, it can be difficult for non-expert system designers
to choose the appropriate configuration to use.
Moreover, these mechanisms are generally configured once for all, which results in the same configuration for every protected piece of information.
However, not all users have the same behaviour, and even the behaviour of a single user is likely to change over time.
To address this issue, we present in this paper \accio, a new framework enabling the dynamic configuration of LPPMs.
\accio can be used in two scenarios: (1) \textit{offline}, where \accio enables a system designer to choose and automatically tune the most appropriate LPPM for the protection of a given dataset; (2) \textit{online}, where \accio enables the user of a crowd sensing application to protect consecutive batches of her geolocated data by automatically tuning an existing LPPM to fulfil a set of privacy and utility objectives.
We evaluate \accio on both scenarios with two real-life mobility datasets and two state-of-the-art LPPMs.
Our experiments show that the adaptive LPPM configurations found by \accio{} outperform both in terms of privacy and utility a set of static configurations manually fixed by a system designer.
\end{abstract}

\section{Introduction}\label{sec:intro}

More and more users are equipped with GPS-enabled handheld devices, such as smartphones, tablets or smart watches, on which they run geolocated applications.
Examples of such applications include crowd sensing services such as Waze\footnote{\url{https://www.waze.com/}}, which provides real-time traffic information to the users or geolocated social networks such as Swarm\footnote{\url{https://www.swarmapp.com/}} (formerly Foursquare), which turns location sharing into a game.
Whatever their exact nature, most of these \textit{location-based services} (LBSs) have the same goals: on the one side, they use geolocated data generated by users to provide them a free, accurate and contextual service and on the other side they make business out of the collected data.
Indeed, mobility data is highly valuable and the market related to LBSs is huge: total annual revenue of the US-only LBSs industry was already estimated to \$75 billions in 2012~\cite{Henttu12}.

The downside of the picture is that this increasing use of LBSs inevitably leads to a huge and constantly increasing amount of mobility data being collected and leaked everyday about individuals. 
Indeed, analysing mobility traces of users can reveal their \textit{points of interest}~\cite{Gambs11} (POIs, which are meaningful places such as their homes or work places), the other users they frequently meet~\cite{Sharad14}, or lead to predicting their future mobility~\cite{Sadilek12}.
It is also possible to semantically label these mobility traces in order to infer even more 
sensitive information (e.g., sexual, religious or political preferences if one regularly goes to a gay bar, a worship place or the headquarters of a political party)~\cite{Franceschi15}.

To address the challenge of location privacy, many location privacy protection mechanisms (LPPMs) have been recently proposed~\cite{Acs14,Andres13,Mapomme15b,Fawaz14}.
Some of these LPPMs (e.g., \cite{Acs14,Andres13,Mapomme15b}) can be used \textit{offline} by a system designer to protect a dataset before releasing it while others can also be used \textit{online} by an end-user (e.g., \cite{Andres13,Mapomme15b,Fawaz14}) to obfuscate a data portion before sending it to a given LBS.
However, in both cases the effectiveness of these solutions usually rely on the tuning of a set of configuration parameters (possibly with a large range of possible values), which is a difficult task for non-expert users or system designers as these parameters have both an impact on the privacy offered to the users and on the utility of the protected data.
Moreover, most of the time these parameters are statically set up once and for all, and do not dynamically evolve according to the content of the data under analysis, especially in the online scenario.
Such static LPPM parametrisation may however lead to the over protection of non sensitive data portions (e.g., a portion of the data without any point of interest) 
thus uselessly degrading its utility, and to the under protection of possibly sensitive data portions (e.g., the regular visit of a hospital), thus resulting in the leakage of sensitive information about the user.

A few frameworks for evaluating and comparing LPPMs~\cite{Shokri11} as well as adaptive LPPMs~\cite{Agir14,Chatzikokolakis15} have been presented in the literature.
However, these works put the emphasis on privacy guarantees offered to users, but are rarely interested in the utility of the resulting data.

In this paper, we present \accio (which stands for Adaptive Location Privacy), a new framework for evaluating and dynamically configuring LPPMs, which considers both privacy and utility as equally important objectives.
Specifically, \accio contains a generic model enabling the specification of a set of privacy and utility \textit{objectives} that the LPPM shall satisfy.
Then, instead of testing static configuration parameters for each LPPM, \accio uses an \textit{optimizer} that dynamically tunes the parameters of the LPPM under consideration according to the current data portion to which it is applied on in order to meet the privacy and utility objectives specified by the system designer. 

The generality of \accio allows its deployment both offline for comparing and tuning a set of LPPMs with the purpose of protecting a static dataset before releasing it and online in the context of a crowd sensing application for dynamically configuring a given LPPM with respect to the given data portion under analysis.
In both cases, the major contribution of \accio is its ability to automatically find LPPM configurations that fulfil a set of possibly conflicting privacy and utility objectives that it would be cumbersome to find manually otherwise.

We illustrate the capabilities of \accio{} by comparing two state-of-the-art LPPMs, i.e., Geo-Indistinguishability~\cite{Andres13} (\geoi for short), which applies spatial distortion to the mobility data and \promesse~\cite{Mapomme15b}, which applies temporal distortion to the mobility data, on two real mobility datasets involving 182 and 185 users and containing 25M and 11M location records, respectively.
We show in an offline scenario that \accio eases the comparison of these LPPMs by relying on a set of metrics provided by the framework. 
We further show in an online scenario where users periodically send batches of their data to a third party LBS, that \accio is able to dynamically find configurations of these LPPMs that outperform a set representative static configurations of used LPPMs.
For instance, we show that \accio is able to tune \geoi on a per batch basis enabling the perfect hiding of points of interest for at least 75~\% of the batches while having a spacial distortion lower than 150m on the two datasets.
The results for \promesse are even better as \accio is able to find for each batch a configuration of the protocol enabling to globally outperform all the representative static configurations both on the considered privacy and utility metrics, thus reaching the best of the two worlds.
To assess the usability of \accio on mobile devices we measured the latency of running \accio on an emulated Android smartphone.
Results show that the latency of running \accio is highly dependent on the LPPM under consideration, with an average execution time of 9s with \geoi and 500ms with \promesse.
Finally, for enabling the reproducibility of our results and the reuse of our framework, the code of \accio and the used datasets are publicly available~\cite{privamov-srds16}.


The remaining of this paper is structured as follows.
We first review the related works in Section~\ref{sec:related}.
We present an overview of \accio in Section~\ref{sec:overview} followed by the system model in Section~\ref{sec:models}.
The privacy and utility evaluation and optimization processes are detailed in sections~\ref{sec:metrics} and~\ref{sec:optimizer}, respectively.
Finally, we present our experimental evaluation in Section~\ref{sec:eval} and conclude this paper in Section~\ref{sec:conclusion}.

\section{Related work}\label{sec:related}

\emph{Location privacy protection mechanisms} (LPPMs) attempt to protect users against privacy threats by obfuscating their mobility traces before sending them to third party LBSs.
The two most adopted privacy guarantees provided by LPPMs follow the $k$-anonymity~\cite{Sweeney02} 
and the $\epsilon$-differential privacy~\cite{Dwork06} models.
While the former hides a user within cloaking areas containing at least $k-1$ other users, the latter disturbs the mobility traces in such a way that it theoretically bounds by a factor $\epsilon$ the impact of the removal of a single element of the dataset.
For instance, \cite{Mokbel06} describes a protection mechanism providing $k$-anonymity that relies on a centralized anonymisation proxy.
In this protocol, the proxy receives all user queries and generates cloaking areas before sending the obfuscated query to the LBS.
It then extracts the response before returning it to the user.
In~\cite{Ghinita07}, the authors removed the dependency to a trusted proxy by presenting a distributed protection mechanism that dynamically builds cloaking areas of at least $k$ users during their mobility.
More recently, Geo-Indistinguishability (\geoi for short)~\cite{Andres13} was proposed as an extension of differential privacy to be used on mobility data.
The guarantee is enforced by adding a calibrated noise drawn from a two-dimensional Laplace distribution.
\geoi has been successfully applied in an online context when a user is querying an LBS in real-time, and in an offline context when an entire dataset gathering the mobility traces of a set of users is protected to be released.
\promesse~\cite{Mapomme15b} is another protection mechanism whose goal is to hide POIs from traces via speed smoothing.
More precisely, instead of obfuscating locations, \promesse obfuscates the temporal dimension of traces.
This approach introduces almost no spatial error to the obfuscated data, at the cost of a reduced temporal accuracy.

While all the proposed approaches try to improve the degree of protection offered to the users, none of them consider the utility (or accuracy) of the protected mobility traces as an input parameter of the protection mechanism.
In practice, there is an inherent trade-off between utility and privacy.
For instance, \cite{Krumm07} showed that the amount of noise required to include in the raw traces in order to defeat the best-performing known attacks would likely make protected traces unusable by any LBS.
It is thus important to consider utility in the balance by taking into account the usefulness of the resulting protected mobility traces.

\textbf{Configuring LPPMs.}
The effectiveness of obfuscation mechanisms also relies on the appropriate setting of a set of configuration parameters, which can be difficult for non-expert system designers.
For instance, the $\epsilon$ parameter of an $\epsilon$-differentially private LPPM must be defined and has a great impact on the offered protection, although it can be difficult to have an a priori feeling of the exact difference between two real values of this parameter.
In addition, most of LPPMs are statically configured once and for all, regardless of the evolution of the incoming mobility traces to obfuscate.
Indeed, as the behaviour of users changes over time, the properties of their mobility data change accordingly (e.g. speed, sampling rate, places visited).
Hence, while a given LPPM configuration can be effective at a given time period it may become ineffective at another time period.
Few initiatives have been proposed to dynamically adapt the protection mechanisms according to the underlying data.
Chatzikokolakis et. al~\cite{Chatzikokolakis15} proposed an extension of \geoi, which leverages contextual information (i.e., if the user is located in an urban environment or a countryside area) to calibrate the amount of noise applied to disturb the mobility traces.
Agir et. al~\cite{Agir14}, in turn, introduced an adaptive mechanism to dynamically change the size of an obfuscated area hiding the exact location of users.
More precisely, the proposed solution locally evaluates the privacy level and enlarges the area accordingly until a target privacy level is achieved.
However, these approaches are designed with a single privacy goal in mind and do not give utility the same level of importance.

\textbf{Evaluating LPPMs.}
Evaluating existing LPPMs has been at the center of a few initiatives. Among the difficult challenges in this context is the heterogeneity of privacy and utility metrics.
Indeed, not all LPPMs reported in the literature use the same privacy and utility metrics, which makes their comparison difficult.
As described in Section~\ref{sec:metrics}, \accio implements multiple privacy and utility metrics, which makes the comparison of various LPPMs easier.
Finally, similarly to \accio, \cite{Shokri11} proposes a full location privacy framework.
To evaluate the performance of an LPPM, this solution compares the outcome of a privacy attack performed on a raw trace against the same attack performed on its obfuscated counterpart.
However, this solution only works for probabilistic LPPMs and is not adapted to more generic mechanisms (e.g., \promesse~\cite{Mapomme15b}).
Moreover, only the privacy measurement is evaluated and the trade-off between privacy and utility is not considered.

\section{\accio overview}\label{sec:overview}

\begin{figure}
	\centerline{\includegraphics[width=0.5\textwidth]{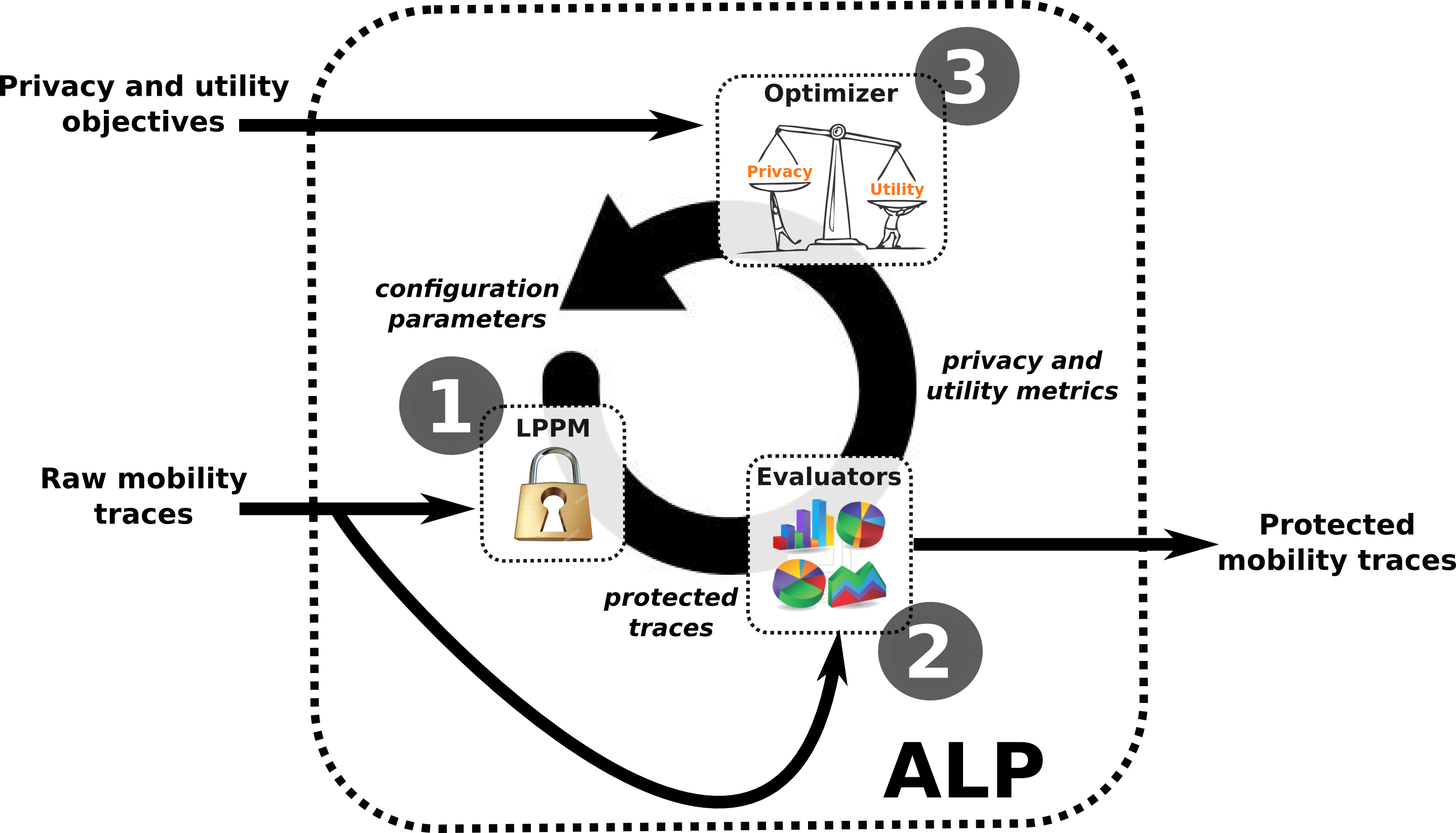}}
	\caption{Components forming the \accio framework. Mobility traces are read from a mobile device (online scenario) or a local dataset (offline scenario). The protection mechanism is ran and its output evaluated with respect to some user-defined objectives. An optimizer then analyses the metrics and iteratively tries to propose a better configuration.}
	\label{fig:overview}
	\vspace{-10pt}
\end{figure}

We present in this section an overview on \accio{}, a framework for the dynamic configuration of LPPMs. As depicted in Figure~\ref{fig:overview}, \accio{} takes as input raw mobility traces and outputs protected mobility traces. However, contrary to existing LPPMs, \accio{} does so by also considering a set of privacy and utility objectives specified by the data holder (the user or the system designer). 
These objectives are specified in \accio{} using a model further defined in Section~\ref{sec:metrics}.
To protect raw traces, \accio{} proceeds as follows.
First, the raw traces get obfuscated using a given LPPM applied with an initial (random) configuration (step 1 in the figure).
The protected traces are then evaluated with respect to the specified privacy and utility objectives (step 2 in the figure).
Then an optimisation process uses the result of this evaluation to iteratively propose a better configuration for the LPPM (step 3 in the figure).
In \accio, this optimisation process, which is further presented in Section~\ref{sec:optimizer}, is based on the simulated annealing algorithm~\cite{Kirkpatrick82}.
This step outputs new values for the LPPM configuration parameters, which are re-used in another round of step 1.
The three steps are repeated until a satisfactory configuration is found. 

\begin{figure*}[!t]
\centering
\subfloat[\accio offline scenario]{\includegraphics[width=8cm]{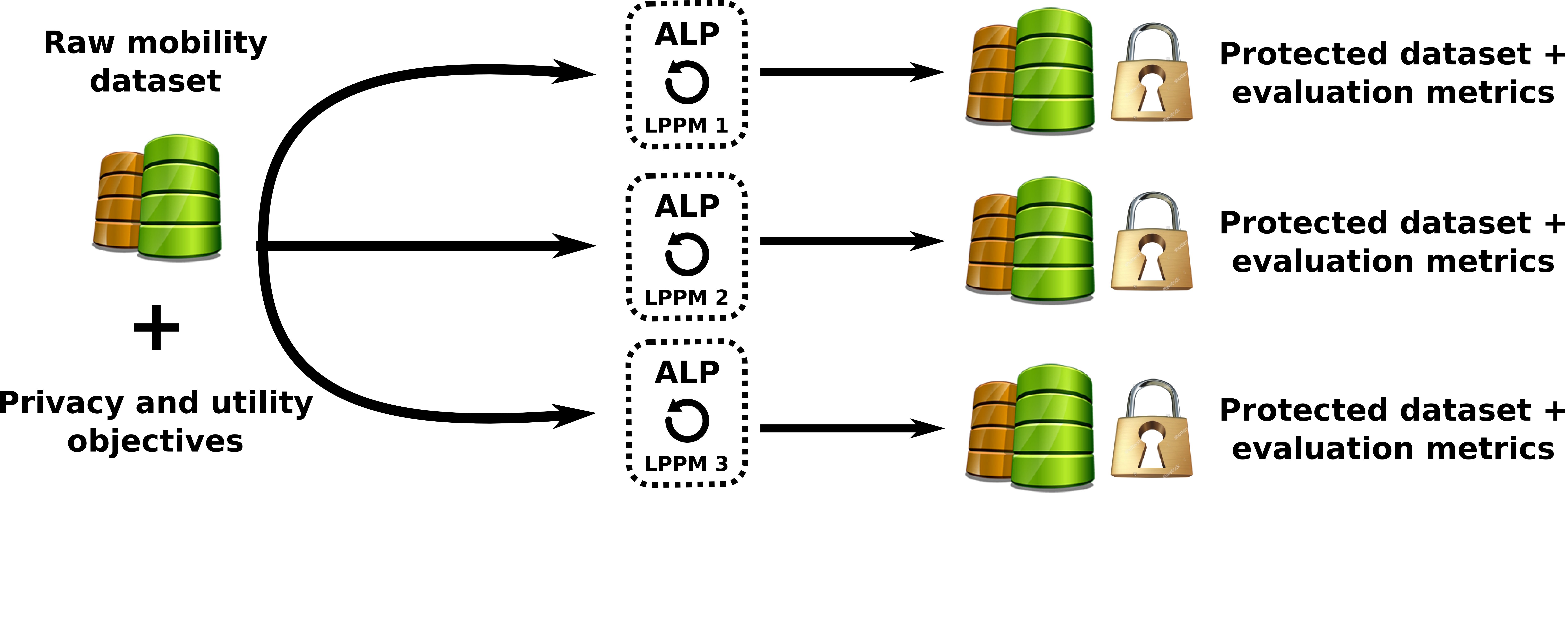}\label{fig:offline}}
\hfil
\subfloat[\accio online scenario]{\includegraphics[width=8cm]{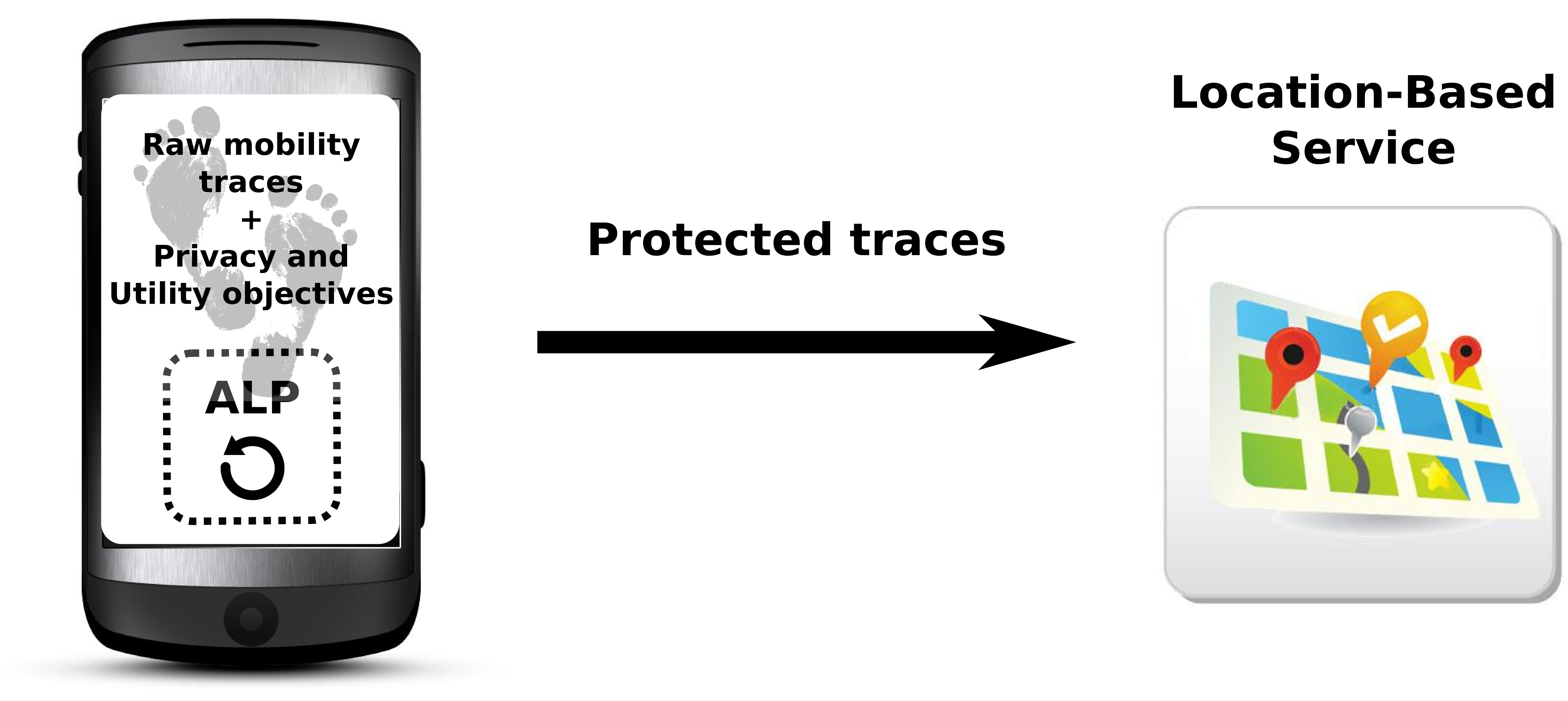}\label{fig:online}}
\caption{\accio in action: the offline protection of a complete dataset before releasing it (left) and the online optimisation of an LPPM
for individual users periodically interacting with an LBS (right).}
\label{fig:overview_outside}
\vspace{-0.45cm}
\end{figure*}

As depicted in Figure~\ref{fig:overview_outside}, \accio can be used in two major scenarios. 
The first scenario is the offline protection of a complete dataset (Figure~\ref{fig:offline}). 
In this scenario, a system designer wants to protect a dataset of mobility traces before releasing it. Towards this purpose, he uses \accio to automatically tune different LPPMs according to a set of privacy and utility objectives he would like to achieve. 
As a result, the system designer gets the result of a set of evaluation metrics for each configured LPPM, which allows him to decide which corresponding obfuscated dataset to release.

The second scenario is the online optimisation of an LPPM for individual users periodically interacting with an LBS (Figure~\ref{fig:online}). This could be the case of a crowd sensing application.
In this scenario, \accio is deployed on the mobile device of a user to protect his mobility data, which is periodically produced and sent to the crowd sensing server.
To achieve that, \accio dynamically tunes an LPPM according both to a set of privacy and utility objectives set by the user and to the current data under analysis.

In both scenarios, the key feature of \accio is its ability to dynamically optimize an LPPM with respect to a set of privacy and utility objectives.

\section{System model}\label{sec:models}

This section presents a set of notations that will be instrumental for introducing the contributions of this paper.



We assume that users are identified through \textit{user identifiers}, which are elements of $\mathcal{U}$.
A \textit{location} is a point on the surface of the Earth.
There are several ways to represent locations, such as a latitude-longitude pair or a projection in Cartesian coordinates.
We abstract this by considering locations as elements of $\mathcal{L}$ with an associated distance function $d_\mathcal{X} : \mathcal{L}^2 \to \mathbb{R}^+$.
Locations are generally associated with a \textit{timestamp}, which is an absolute instant (it does not convey timezone information). Timestamps are elements of $\Omega$ and have a partial order.
Note that this means that we stay generic and consider location and time as continuous.
A \textit{record} is defined as the location of a user at a specific timestamp. More specifically, a record is a triplet $\langle u, \ell, t \rangle \in \mathcal{R}$, where $u \in \mathcal{U}$, $\ell \in \mathcal{L}$, $t \in \Omega$.
A \textit{trace} of user $u$ is a vector of chronologically ordered records belonging to $u$.
The set of all possible traces is noted $\mathcal{T}$.
A \textit{dataset} is defined as a set of traces.
Inside a dataset, there may be one or many traces associated with a single user identifier.



%

\section{Evaluating privacy \& utility in \accio{}}
\label{sec:metrics}

Key to the evaluation of a protection mechanism is the definition of well-suited privacy and utility metrics. These metrics are then practically computed in \accio using metric evaluators implemented as part of the framework.
The list of all available metrics defined in \accio is summarised in Table~\ref{tab:evaluators}.
These metrics can be used to measure either the privacy or the utility. 
This section describes these metrics and gives more insight to the ones used in the evaluation of \accio{} (Section~\ref{sec:eval}).
These metrics will ultimately be available to the final user to allow him to specify his objectives in terms of privacy and utility.

\begin{table}[!ht]
\renewcommand{\arraystretch}{1.3}
\caption{Evaluation metrics available in \accio{}}
\label{tab:evaluators}
\centering
\begin{tabular}{l|c|c}
\hline
\textbf{Metric evaluator} & \textbf{Domain} & \textbf{Purpose}\\
\hline
POIs retrieval & $\mathbb{R}^+$ & Privacy\\
\hline
Spatial distortion & Distance & Utility\\
\hline
Area coverage & $\mathbb{R}^+$ & Utility\\
\hline
\end{tabular}
\vspace{-3mm}
\end{table}

\subsection{POIs retrieval}

Points of interest (POIs) are a very sensitive piece of information, allowing to capture locations that do matter for the users like their home or work places.
Consequently, the number of POIs that can be extracted from a trace is generally considered as a privacy metric~\cite{Mapomme14}, which one would like to minimise, as we first proposed in~\cite{Mapomme15b}.
More specifically, we use the F-Score metric to quantify both the proportion of POIs that are successfully and wrongfully inferred from the obfuscated trace.
The proportion of successfully inferred POIs (i.e., the recall, formally defined in Definition~\ref{def:recall}) gives a hint about the power of the attacker. 
The proportion of wrongfully inferred POIs (i.e., the opposite of the precision, formally defined in Definition~\ref{def:precision}) gives a hint about the confusion of the attacker about whether a POI is real or not.
POIs can be extracted using a simple spatio-temporal clustering algorithm parametrised with a maximum POI diameter $\Delta \ell$ and a minimum stay time $\Delta t$, such as the one proposed in~\cite{Hariharan04}.
Specifically, to measure the privacy leakage, we extract POIs from the dataset before and after having applied an LPPM and evaluate how closely the latter matches with the former.
In the following definitions, we note $P \in \mathcal{L}^n$ a set of POIs extracted from an original trace and $P' \in \mathcal{L}^{n'}$ a set of POIs extracted from a protected trace. In order to easily compare POIs, we abstract them as simple locations, which are defined as the centroid of the cluster of points in which a user stayed for a parametric period of time.

\begin{definition}\label{def:recall} The recall is the ratio between the number of POIs extracted from the obfuscated trace actually corresponding to an actual POI (within a threshold $\ell$) and the number of POIs extracted from the original trace:
\[recall_\ell(P, P') = \frac{|\{p' \in P' \,|\, \exists p \in P, d_\mathcal{X}(p, p') \leq \ell\}|}{|P|}.\]
\end{definition}

\begin{definition}\label{def:precision} The precision is the ratio between the number of POIs extracted from the obfuscated trace actually corresponding to an actual POI (within a threshold $\ell$) and the number of POIs extracted from the obfuscated trace:
\[precision_\ell(P, P') = \frac{|\{p' \in P' \,|\, \exists p \in P, d_\mathcal{X}(p, p') \leq \ell\}|}{|P'|}.\]
\end{definition}

\begin{definition} The POIs retrieval is the harmonic mean of precision and recall:
\[pois_\ell(P, P') = \frac{2 \times precision_\ell(P,P') \times recall_\ell(P,P')}{precision_\ell(P,P') + recall_\ell(P,P')}.\]
\end{definition}



\subsection{Spatial distortion}

This metric assesses the spatial imprecision between locations before and after the obfuscation process, as we first proposed in~\cite{Mapomme15b}.
The spatial distortion introduced by an LPPM has a direct impact on the utility.
The spatial distortion is a distance, expressed in the same unit as the output of $d_\mathcal{X}$.
In our evaluation, we will consider the Euclidian distance, so it will be expressed in meters.
We note $L \in \mathcal{L}^n$ a set of locations from records of an original trace and $L' \in \mathcal{L}^{n'}$ a set of locations from records of a protected trace.

\begin{definition} The spatial distortion is the average distance between protected locations and the closest real location:
\[distortion(L,L') = \frac{\sum\limits_{\ell' \in L'} \min\limits_{\ell \in L} d_\mathcal{X}(\ell, \ell')}{|L'|}.\]
\end{definition}

\subsection{Area coverage}

LPPMs modify mobility data to protect sensitive information. 
Consequently, protection mechanisms may remove locations considered as too sensitive for the user.
This side effect ultimately results in an alteration of the utility of the protected data and may reduce the quality of service of LBSs.
To take into account this side effect on utility, we consider the area coverage, more precisely the proportion of regions for which data is available in the obfuscated traces out of all the regions covered by the raw traces.
The idea of such a metric has been first proposed in~\cite{Abul08}, we define it formally here.

Similarly to POIs retrieval, we use an F-Score to take into account both the proportion of regions for which there is still data in the obfuscated traces and the proportion of regions for which we wrongfully receive data in these traces.
To measure the utility, we discretise the world into cells of variable size and compare the cells that are represented before and after obfuscation.
In the following definition, we consider that we have a function $cell: \mathcal{L} \rightarrow \mathbb{R}$ that assigns to a location a cell identifier.
We note $L \in \mathcal{L}^n$ a set of locations from records of an original trace and $L' \in \mathcal{L}^{n'}$ a set of locations from record of a protected trace.

\begin{definition} The recall is the ratio between the cells extracted from the obfuscated trace actually corresponding to a cell represented in the original trace and the number of cells in the original trace.
\[recall(L,L') = \frac{|\{cell(\ell') \,|\, \ell' \in L'\} \cap \{cell(\ell) \,|\, \ell \in L\}|}{|L|}.\]
\end{definition}

\begin{definition} The precision is the ratio between the cells extracted from the obfuscated trace actually corresponding to a cell represented in the original trace and the number of cells in the obfuscated trace.
\[precision(L,L') = \frac{|\{cell(\ell') \,|\,  \ell' \in L'\} \cap \{cell(\ell) \,|\, \ell \in L\}|}{|L'|}.\]
\end{definition}

\begin{definition} The area coverage is the harmonic mean of precision and recall:
\[coverage(L,L') = \frac{2 \times precision(L,L') \times recall(L,L')}{precision(L,L') + recall(L,L')}.\]
\end{definition}





\section{Optimizing privacy \& utility in \accio{}}
\label{sec:optimizer}

By combining metric evaluators with an optimizer, \accio is able to tune protection mechanisms to achieve a set of privacy and utility objectives.
More precisely, the optimizer receives as input the values of a privacy and utility evaluation metrics associated to the current mobility data and automatically tunes the configuration parameters of the protection mechanism.
To achieve that, the optimizer relies on an instantiation of the simulated annealing algorithm. 
This section starts with a background on the simulated annealing algorithm (Section~\ref{subsec:algorithm}) followed by the various adaptations necessary for using this algorithm in the context of \accio{}, i.e., the definition of a cost function, the randomisation of the explored space and the cooling schedule described in Sections~\ref{subsec:objectives}, \ref{subsec:randomising} and \ref{subsec:cooling}, respectively.

\subsection{Simulated annealing}\label{subsec:algorithm}

Simulated annealing~\cite{Kirkpatrick82} is a well-known probabilistic optimization technique useful to find an approximation of the global optimum of a function.
Finding the exact global optimum is not guaranteed, but this optimization technique ensures an acceptable local optimum in a reasonable amount of time compared to a brute-force method exploring all possible solutions.
It is especially useful for large (of infinite) search spaces.
It follows the physical analogy of cooling down a metal, where the temperature is gradually decreasing until the state is frozen.
If the cooling takes enough time, atoms can find an optimal placement, i.e., a state associated with minimal energy.
The algorithm is depicted in Algorithm~\ref{alg:simulated_annealing}.
The underlying idea is, from an initial state $s \in \mathcal{S}$ (line 2), to probabilistically decide whether to move to a neighbour state $s'$ (lines 9-11) depending on the current temperature and the cost associated with these states (line 8).
This cost corresponds to the energy of a state in the physical analogy.
This process is repeated several times, with a decreasing temperature until the system reaches a minimal temperature (line 5).

\begin{algorithm}\footnotesize
\caption{Simulated annealing algorithm.}
\label{alg:simulated_annealing}
\begin{algorithmic}[1]
    \Function{SimulatedAnnealing}{$t_0 \in \mathbb{R}$, $t_{min} \in \mathbb{R}$, $\delta t \in \mathbb{R}$}
    \State $s \gets \Call{Initial()}{}$
    \State $c \gets \Call{Cost}{s}$
    \State $t \gets t_0$
    \While{$t \geq t_{min}$}
        \State $s' \gets \Call{Neighbour}{s}$
        \State $c' \gets \Call{Cost}{s'}$
        \State $ap \gets \Call{Probability}{c, c', t}$
        \If {$ap \geq \Call{Random}{0,1}$}
            \State $s \gets s'$
            \State $c \gets c'$
        \EndIf
        \State $t \gets t \times \delta t$
    \EndWhile
    \State \Return{$s$}
    \EndFunction
\end{algorithmic}
\end{algorithm}

As shown in the algorithm, a simulated annealing system needs several functions to be defined:    
an initial state function, producing an initial state $s \in \mathcal{S}$ (line 2); 
a neighbour function $\mathcal{S} \rightarrow \mathcal{S}$ associating each state to a neighbouring state (line 6); 
a cost function $\mathcal{S} \rightarrow \mathbb{R}$ associating a cost to each state (lines 3 and 7); 
an acceptance probability function $\mathbb{R}^2 \times \mathbb{R}^+ \rightarrow [0,1]$ giving the probability to accept the new solution given the cost of the current solution, the cost of the new solution and the current temperature (line 8); 
a cooling schedule, controlling the values taken by the temperature (lines 4, 5 and 12).
These functions must be defined according to the particular usage that is being done of the simulated annealing algorithm.
We propose implementations for them in the next sections.

\subsection{Objectives \& cost}\label{subsec:objectives}

\accio{} supports two objectives: maximizing or minimizing a metric\footnote{Extending this set of objectives by supporting comparison operators (e.g., having a metric less than some value) remains future work.}.
The challenge here is to convert some objectives and associated evaluation metric values into a cost (i.e., a single real number) in a such way that the higher the cost, the worst the solution. 
Each objective contributes to a part of the cost. 
Our cost function is depicted in Algorithm~\ref{alg:cost_function}, where \textsc{Evaluate} (line 4) runs the metric evaluator against the current state and returns a single metric value.
As evaluation metrics can be defined in very different ranges (e.g., a distance will be expressed in meters and take values in $\mathbb{R}^+$, whereas a percentage is restricted to $[0,1]$), we normalize them into $[0,1]$ in order to give to each metric a similar weight (line 5).
To achieve that, we impose to each metric a maximum value which bounds the associated cost, and we scale the metric value accordingly.

\begin{algorithm}\footnotesize
\caption{Cost and acceptance probability functions.}
\label{alg:cost_function}
\begin{algorithmic}[1]
    \Data $O \in \mathbb{P}(\mathcal{O})$ a set of objectives
    \Data $ref \in \mathcal{T}$ the trace being obfuscated
    \Function{Cost}{$s \in \mathcal{S}$}
    \State $c \gets 0$
    \For {$o \in O$}
        \State $v \gets \Call{Evaluate}{o.evaluator, ref, s}$ \Comment{Raw value}
        \State $n \gets \min(v, o.scale) / o.scale$ \Comment{Rescaled value}
        \If {$o.minimise$}
            \State $c \gets c + n$
        \Else
            \State $c \gets c + (1 - n)$
        \EndIf
    \EndFor
    \State \Return{$c$}
    \EndFunction
    \Statex
    \Function{Probability}{$c \in \mathbb{R}$, $c' \in \mathbb{R}$, $t \in \mathbb{R}$}
    \If {$c' < c$}
        \State \Return{$1$}
    \Else
        \State \Return{$1/(1+e^\frac{c' - c}{0.5 \times t \times |O|})$}
    \EndIf
    \EndFunction
\end{algorithmic}
\end{algorithm}

Algorithm~\ref{alg:cost_function} also shows the acceptance probability function.
This standard function defines that the probability to accept a solution with a higher cost decreases with the temperature (lines 14-15), although we always accept a smaller cost (lines 12-13).
The $0.5 \times |O|$ expression is a normalization factor.

\begin{algorithm}\footnotesize
\caption{Initial solution and neighbour functions.}
\label{alg:solutions}
\begin{algorithmic}[1]
    \Data $A \in \mathbb{P}(\mathcal{A})$ a set of applicable parameters
    \Function{Initial()}{}
    \State $s \gets \{\}$
    \For {$a \in A$}
        \State $s[a] \gets \Call{RandomValue}{domain}$
    \EndFor
    \State \Return{$s$}
    \EndFunction
    \Statex
    \Function{Neighbour}{$s \in \mathcal{S}$}
    \State $a \gets \Call{RandomParameter}{s}$
    \State $domain \gets \Call{RestrictByHalf}{domain, s[a]}$
    \State $s[a] \gets \Call{RandomValue}{domain}$
    \State \Return{$s$}
    \EndFunction
\end{algorithmic}
\end{algorithm}

\subsection{Randomising solutions}\label{subsec:randomising}
Another challenge of simulated annealing is the way to explore the space of solutions.
In \accio, solutions (or states) are configuration parameters for the considered LPPM.
Each LPPM can be parametrised by several parameters, defined in different ranges of values, possibly infinite.
For example, a $k$-anonymous LPPM should at least have a $k \in \mathbb{R}^+$ parameter defining the level of anonymity, or a basic LPPM randomly dropping records should have a probability $p \in [0,1]$ to keep each record.
In our framework, we consider parameters as a finite set of possible values.
This simplification allows us to consider in the same manner all parameters, whether they are strings, integers, floats, etc.

The exploration of the space of solutions in \accio is presented in Algorithm~\ref{alg:solutions}.
We note as $\mathcal{A}$ the set of all parameters.
First, the initial state of each parameter is defined randomly (lines 1-5). 
Second, the neighbour state is related to the original state (lines 6-10).
More precisely, the new value corresponds to the previous state with one parameter being changed.
It is changed by restricting its domain by half, centered around the previous value.
For example, if the domain of $a \in \mathcal{A}$ is $[1,2,3,4,5]$ and its current value is $2$, the domain when choosing the new value will be restricted to $[1,3]$.

\subsection{Cooling schedule}\label{subsec:cooling}
Finally, a cooling schedule determines the size of the parameter space effectively explored, and affects the acceptance probability.
We chose a static schedule, from $t_0 = 1$ to $t_{min} = 10^{-5}$, with a cooling rate $\delta t = 0.9$.

\begin{figure*}
\centering
\subfloat[Privacy]{\label{fig:offline_eval-privacy}{
  \includegraphics[width=0.32\textwidth]{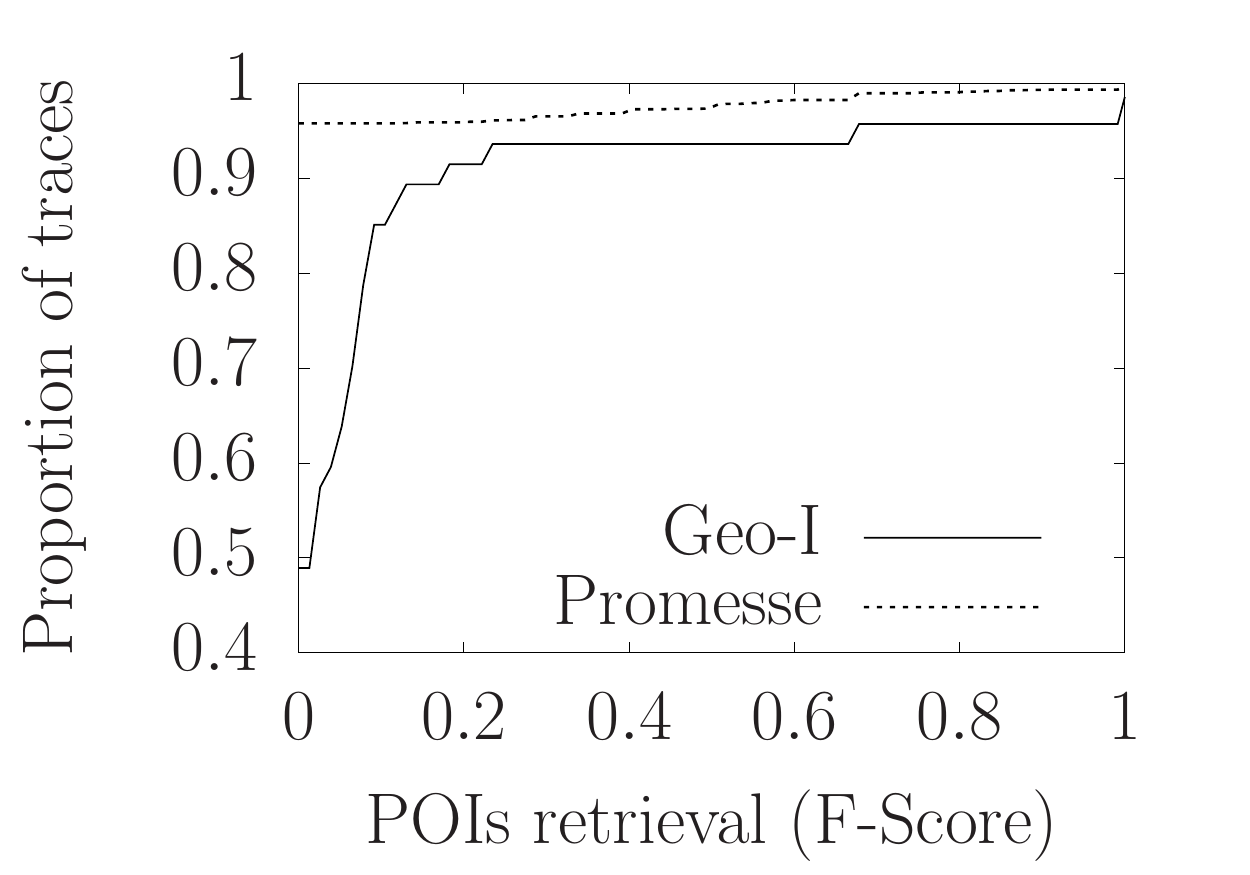}}}
\hfil
\subfloat[Utility -- Spatial distortion]{\label{fig:offline_eval-distortion}{
  \includegraphics[width=0.32\textwidth]{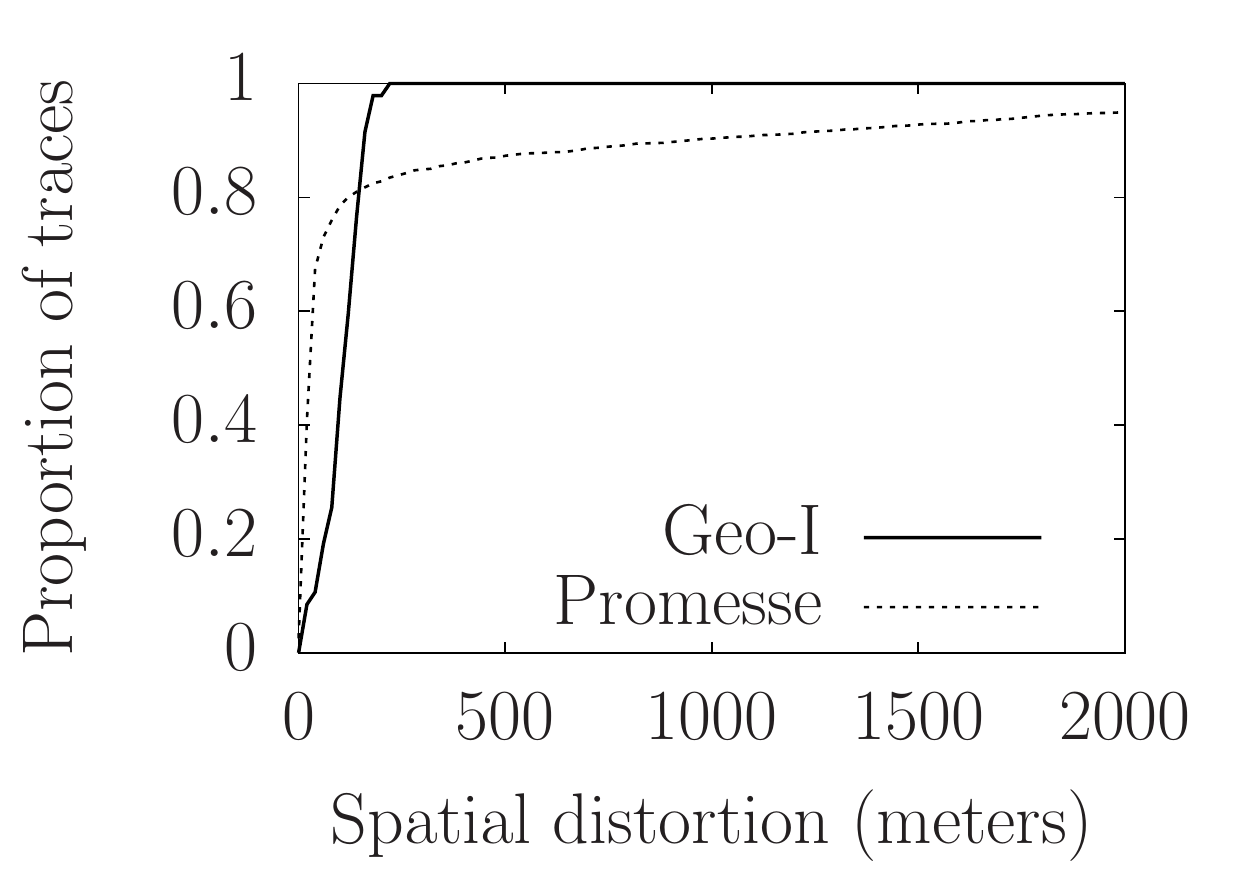}}}
\hfil
\subfloat[Utility -- Area coverage]{\label{fig:offline_eval-coverage}{
  \includegraphics[width=0.32\textwidth]{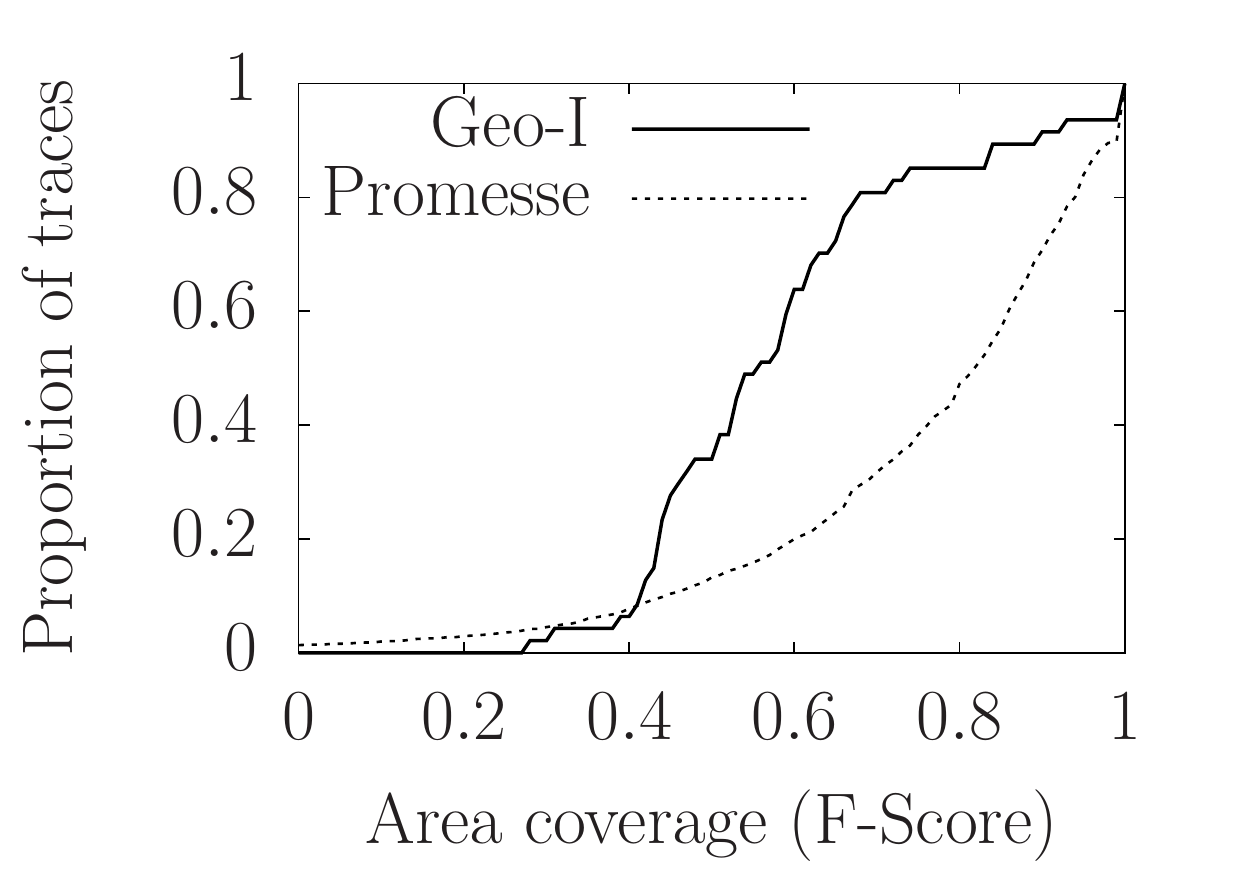}}}
\caption{Cumulative distribution of privacy \& utility metrics with Geolife in the offline scenario.}
\label{fig:offline_eval}
\end{figure*}

\section{Implementation and evaluation}
\label{sec:eval}

This section starts with the presentation of the implementation details of \accio (Section~\ref{subsec:implementation}) and the experimental setup of our evaluation (Section~\ref{sec:settings}).
We then illustrate the capabilities of our framework by evaluating the optimization of two state-of-the-art LPPMs under two different scenarios: (1) an offline scenario where \accio helps a system designer to tune and compare the two LPPMs for obfuscating a whole dataset (Section~\ref{subsec:offline}) and (2) an online scenario where \accio is used by mobile users to fine tune a given LPPM for the obfuscation of batches of geo-located data before sending them to a third party server (Sections~\ref{subsec:tradeoff} and~\ref{subsec:adaptive}).
We finally evaluate the latency of running \accio in a mobile device (Section~\ref{subsec:mobile}).

In a nutshell, our evaluation draws the following conclusions: first, in the offline scenario, the generality of \accio eases the tuning and comparison of state-of-the-art LPPMs.
Further, in the online scenario, \accio allows to find LPPM configurations reaching trade-offs between privacy and utility metrics that outperform representative static configurations of the latter. Finally, the latency of running \accio on a mobile device is reasonable and highly depends on the underlying LPPM.

To enable the reproduction of our experiments, the source code of \accio{} and the used datasets are available on a dedicated website~\cite{privamov-srds16}.

\subsection{Implementation}\label{subsec:implementation}

\accio is implemented in Scala, a language running on the JVM and hence largely interoperable with Java.
\accio is released under an open source license and is publicly available~\cite{privamov-srds16}.
Our location privacy framework is mainly split in two parts.
The first one is a library of common data structures to represent and manipulate mobility data and implementation of state-of-the-art protection mechanisms.
The second part is the glue assembling pieces together and creating the framework.
\accio includes a configuration layer, an optimizer and an execution engine scheduling and running the different operations.
A web-based user interface is also available to easily visualise the output of experiments.
\accio is designed to be extensible and allows researchers as well as practitioners to easily implement their own LPPMs and metric evaluators.

\begin{figure*}
\centering
\subfloat[Privacy -- Geolife dataset]{\label{fig:geoind_eval-pgeo}{
  \includegraphics[width=0.35\textwidth]{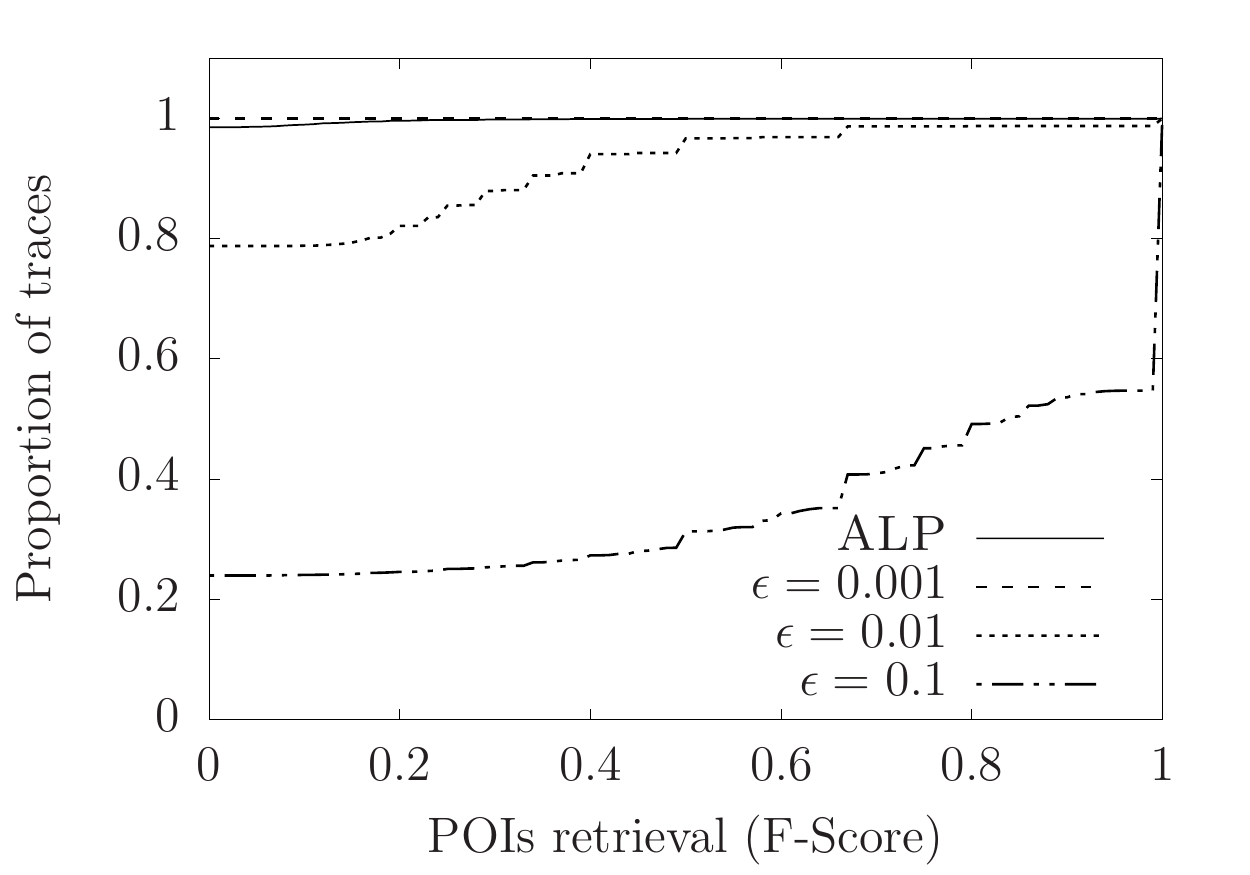}}}
\hfil
\subfloat[Privacy -- MDC dataset]{\label{fig:geoind_eval-pmdc}{
  \includegraphics[width=0.35\textwidth]{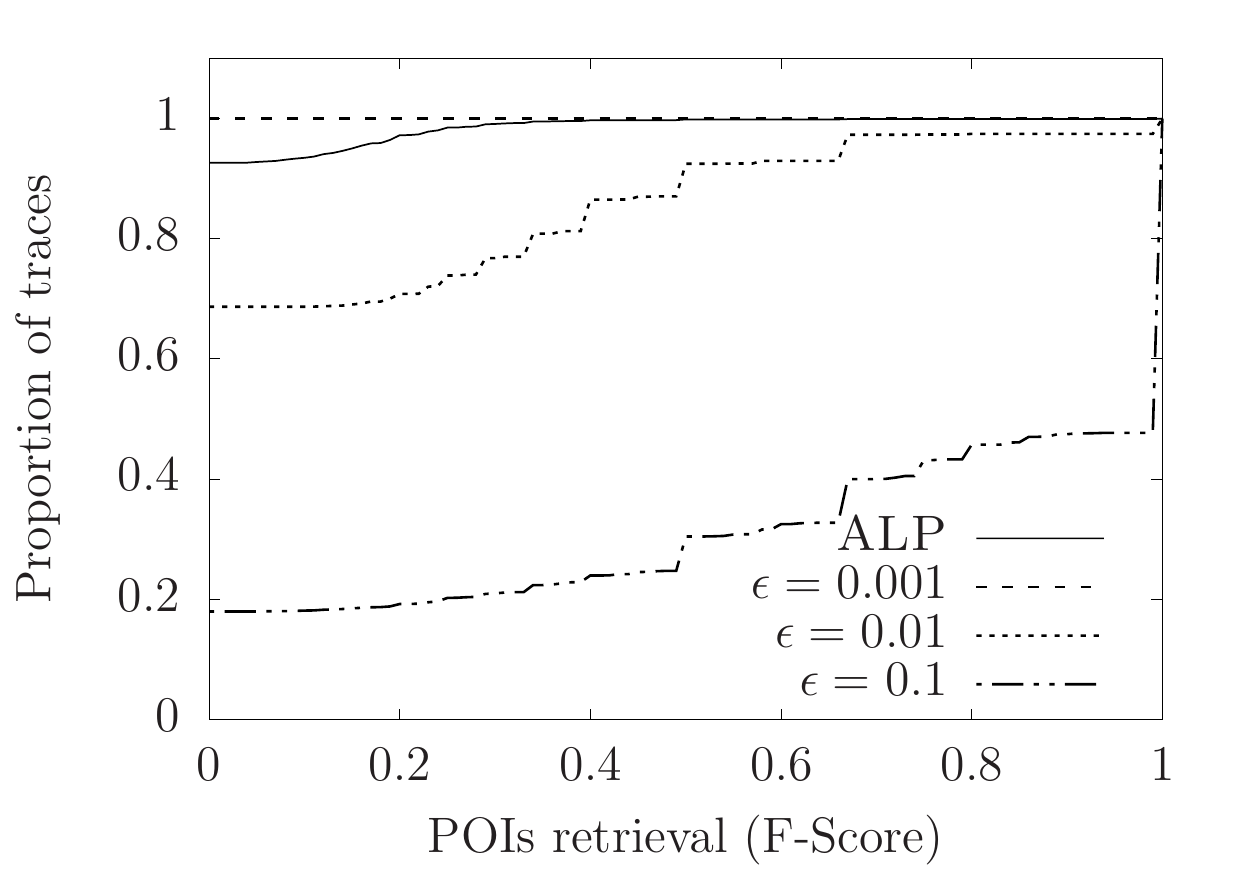}}}

\subfloat[Utility -- Geolife dataset]{\label{fig:geoind_eval-ugeo}{
  \includegraphics[width=0.35\textwidth]{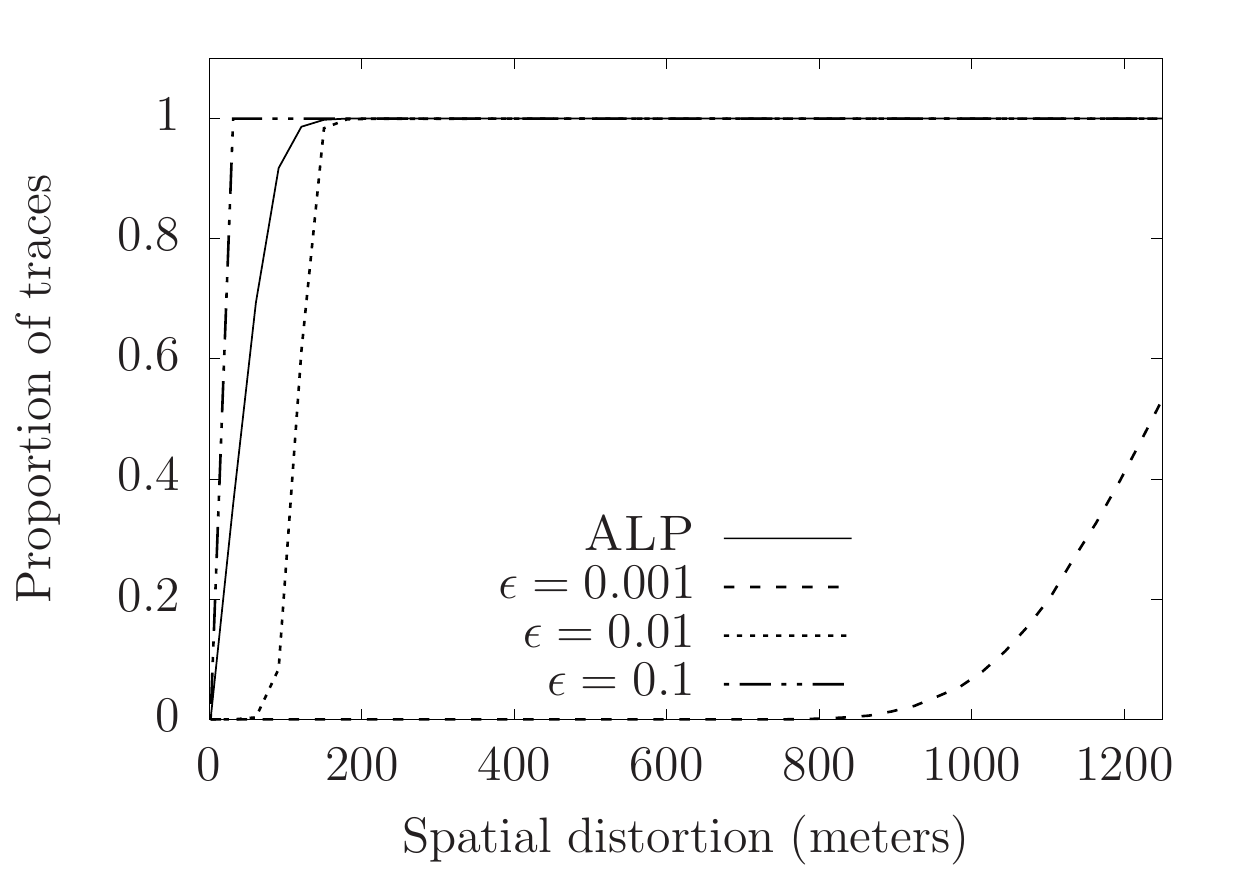}}}
\hfil
\subfloat[Utility -- MDC dataset]{\label{fig:geoind_eval-umdc}{
  \includegraphics[width=0.35\textwidth]{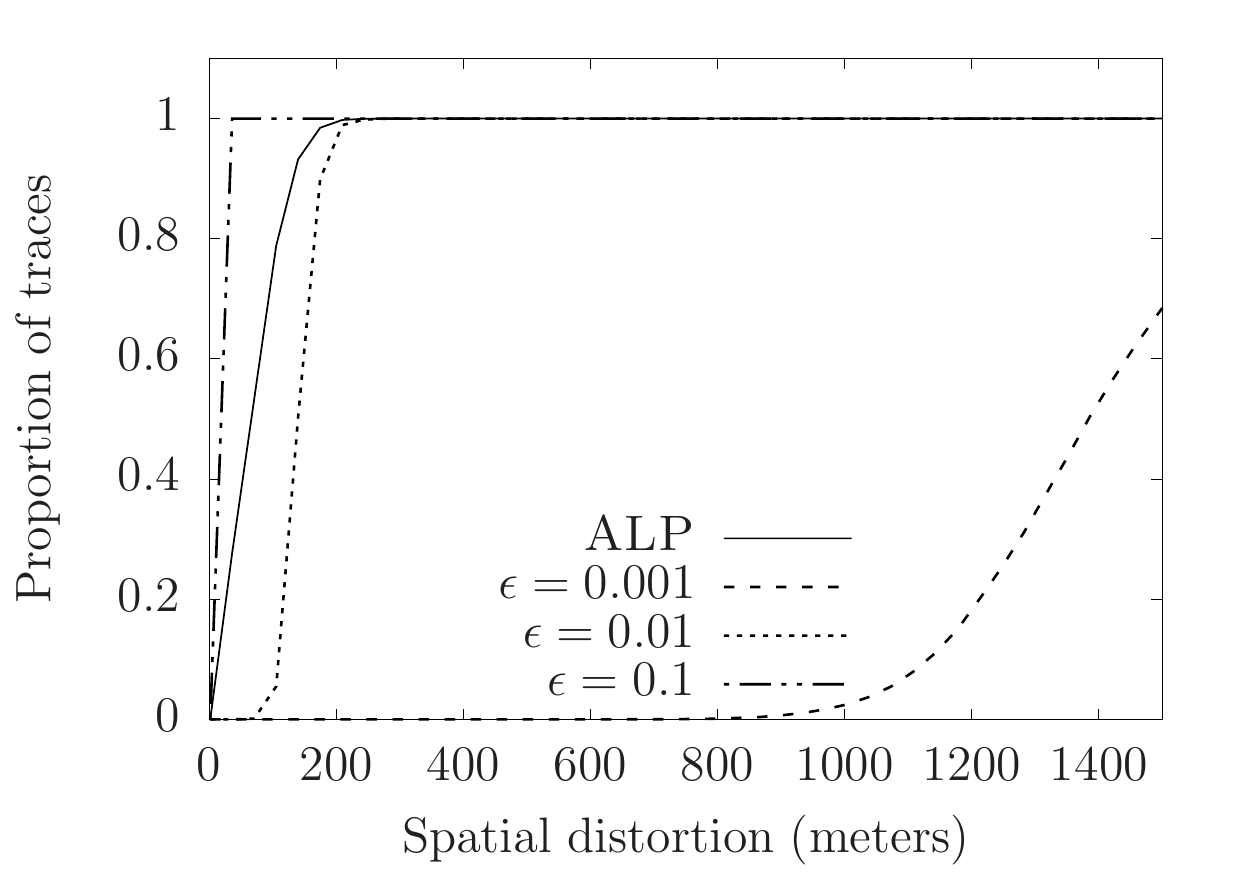}}}
\caption{Cumulative distribution of privacy \& utility metrics under \geoi in the online scenario.}
\label{fig:geoind_eval}
\end{figure*}

\begin{figure*}
\centering
\subfloat[Privacy -- Geolife dataset]{\label{fig:pro_eval-pgeo}{
  \includegraphics[width=0.35\textwidth]{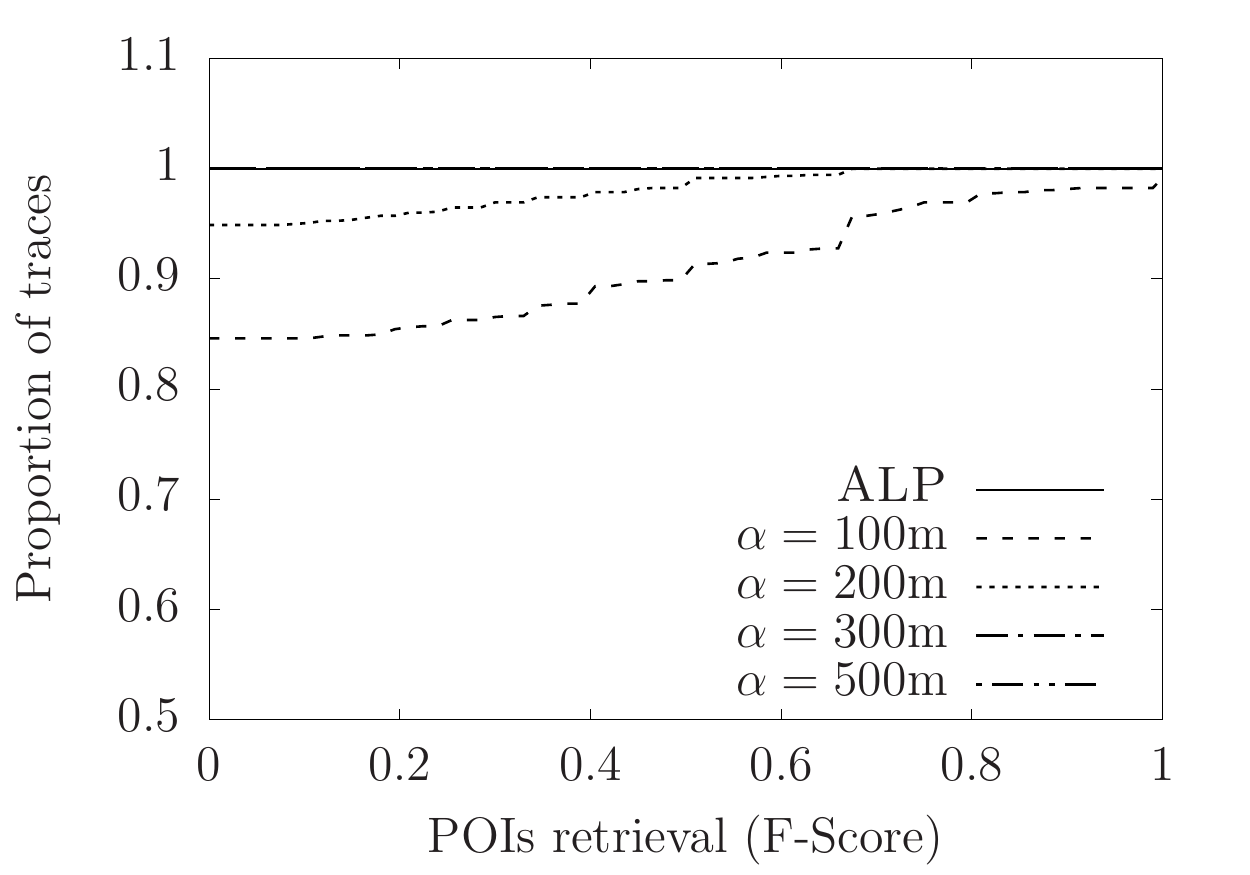}}}
\hfil
\subfloat[Privacy -- MDC dataset]{\label{fig:pro_eval-pmdc}{
  \includegraphics[width=0.35\textwidth]{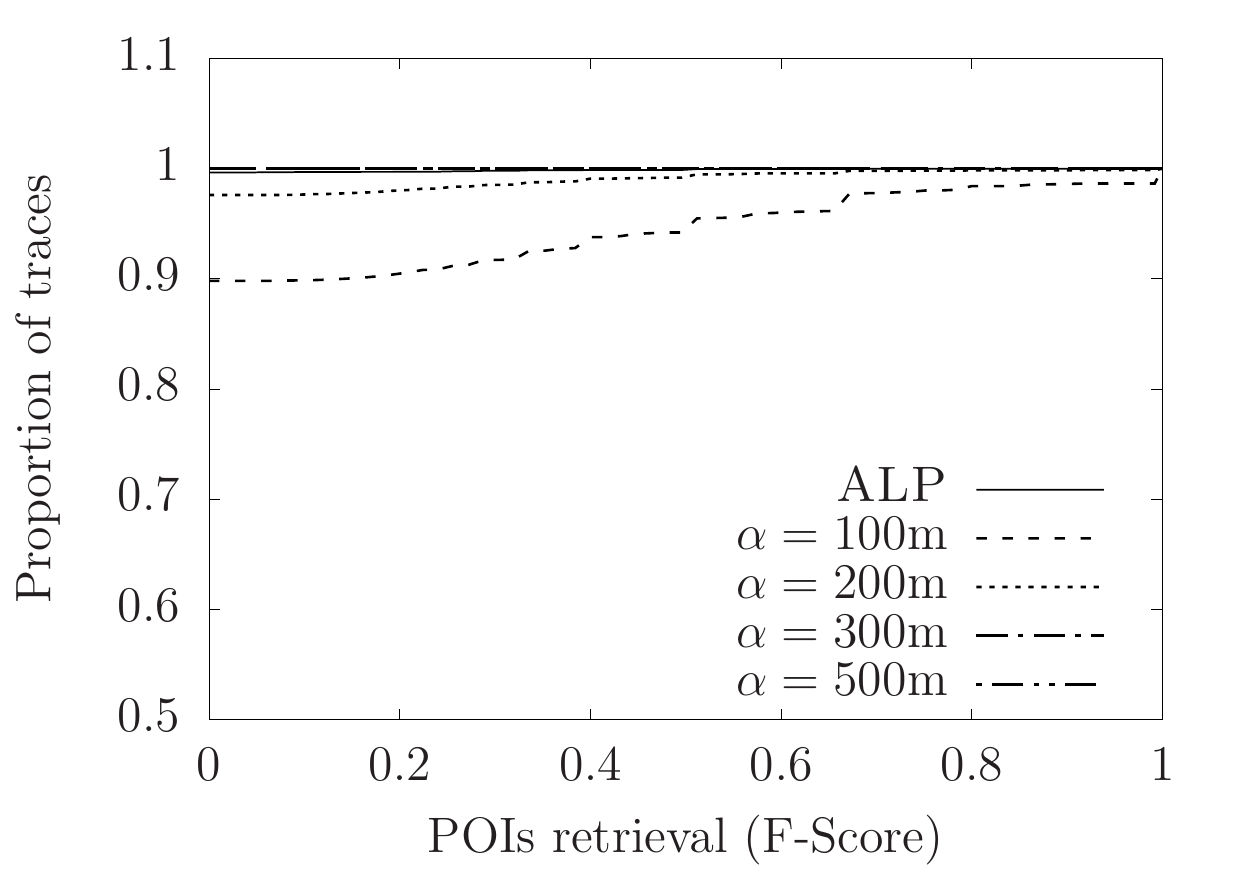}}}

\subfloat[Utility -- Geolife dataset]{\label{fig:pro_eval-ugeo}{
  \includegraphics[width=0.35\textwidth]{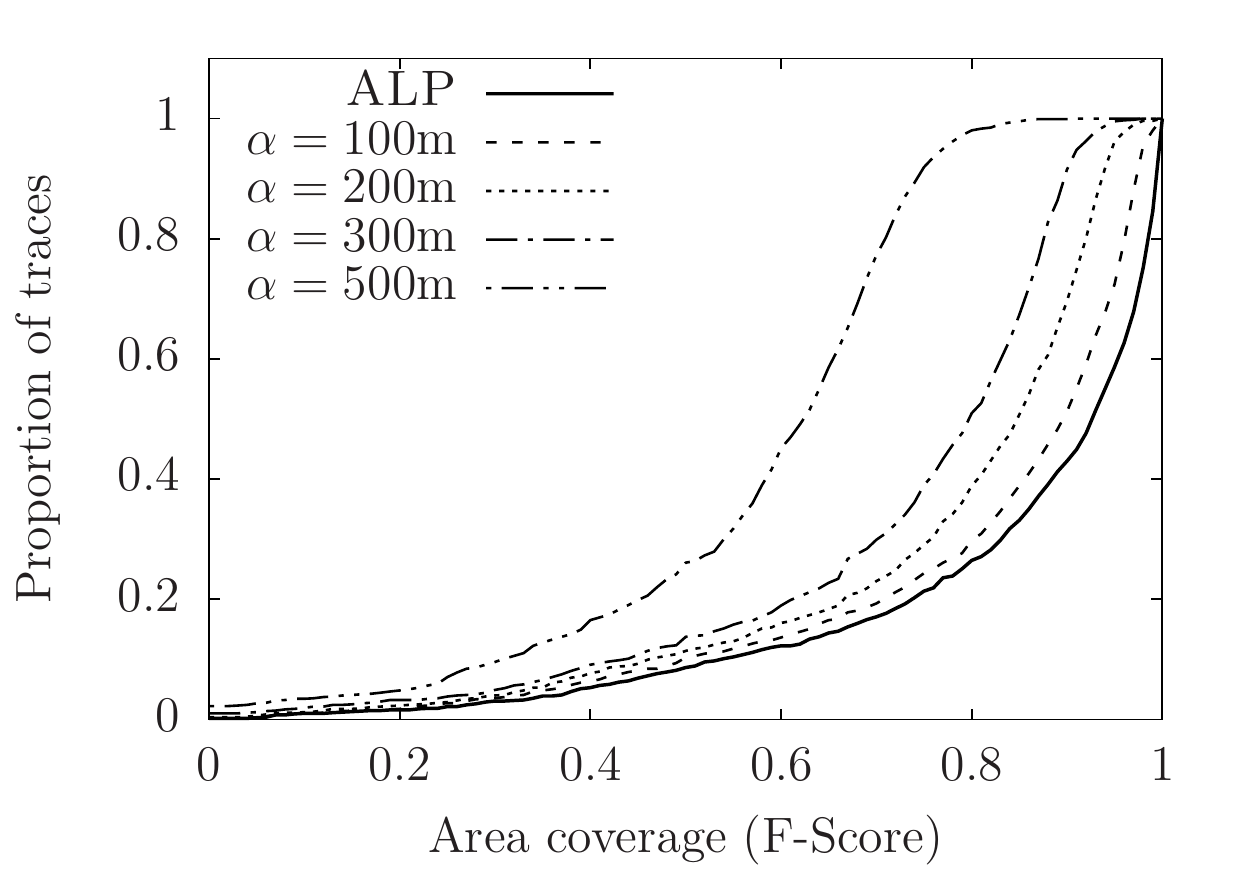}}}
\hfil
\subfloat[Utility -- MDC dataset]{\label{fig:pro_eval-umdc}{
  \includegraphics[width=0.35\textwidth]{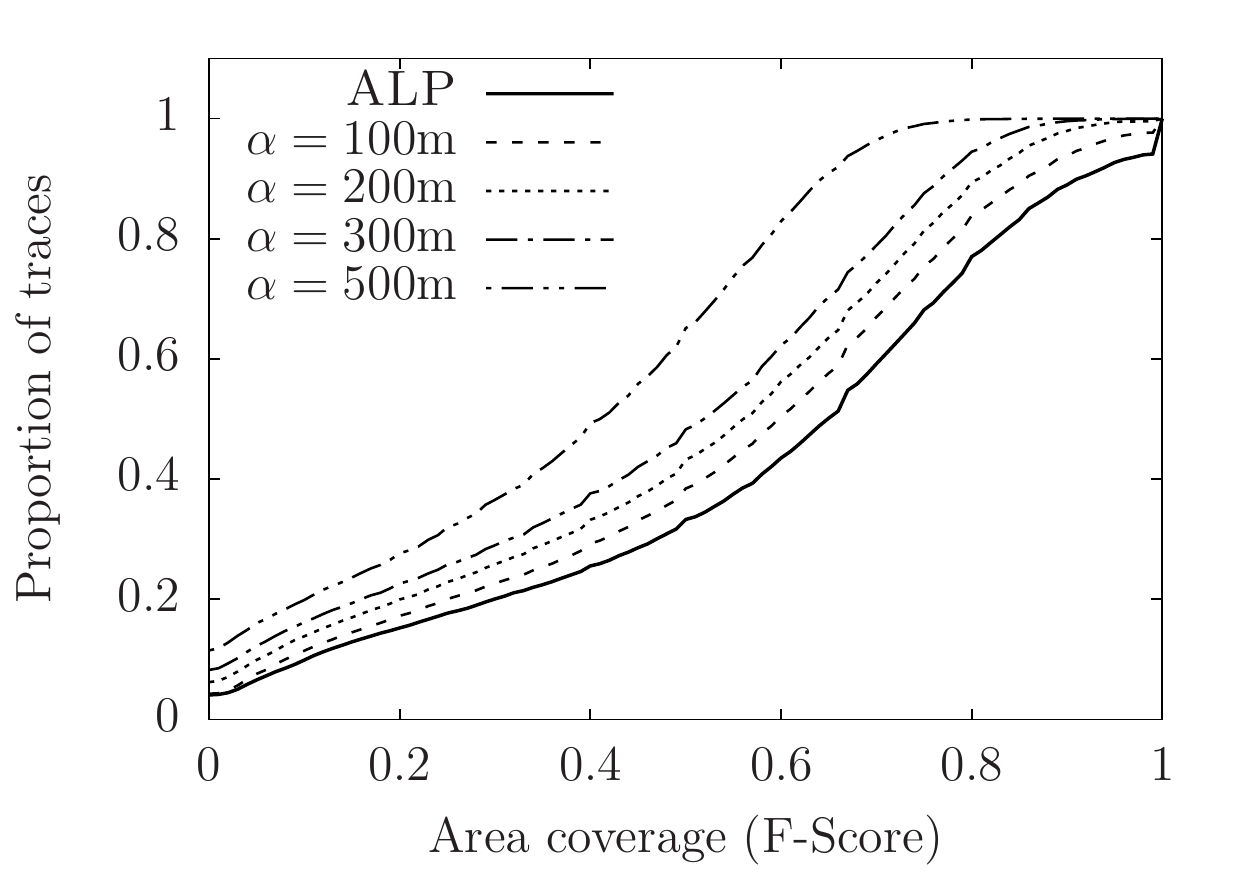}}}
\caption{Cumulative distribution of privacy \& utility metrics under \promesse in the online scenario.}
\label{fig:promesse_eval}
\end{figure*}

\subsection{Experimental setup}
\label{sec:settings}

\textbf{Datasets.}
We evaluate \accio with two real-life datasets: \textit{Geolife}~\cite{Zheng10} and \textit{MDC}~\cite{Kiukkonen10}. 
Geolife has been collected by Microsoft Research Asia over four years and by 182 users.
It includes 25M records, but contains irregularities in term of users activity: some people have been tracked during the whole four years whereas others have only contributed for a few hours.
MDC, in turn, has been collected between 2009 and 2011 around Lausanne, Switzerland and involves a total of 185 users, followed during their daily life.
The entire dataset is composed of 11M records with location information.

\textbf{Protection mechanisms \& objectives.}
We illustrate the capability of \accio through the optimization of two different protection mechanisms: \geoi and \promesse.

\geoi~\cite{Andres13} takes an $\epsilon$ parameter (expressed in meters$^{-1}$) determining the amount of noise to add 
(the smaller $\epsilon$, the higher the amount of noise added to the raw data).
In \accio, $\epsilon$ has been configured to take values in $[0.001, 0.1]$.
Moreover, we use a logarithmic space (in base 10) to draw values for $\epsilon$, because the smallest its value is, the more impact it has on privacy (and therefore utility).
To compare our adaptive solution with statically configured mechanisms, we take as baselines $\epsilon \in \{0.001,0.01,0.1\}$; $0.001$ and $0.1$ and the extreme values that are considered by \accio and $0.01$ gives us a logarithmic progression.
We set as objectives for the optimizer to minimise the POIs retrieval (privacy metric) and to minimise the spatial distortion (utility metric).
We configure the POIs retrieval metric to extract POIs with a 
maximum diameter $\Delta \ell = 200$ meters and a minimum stay time $\Delta t = 15$ minutes.
We use a threshold $\ell = \Delta \ell / 2 = 100$ meters to determine 
whether POIs are correctly retrieved.
Because \geoi is a non-deterministic protection mechanism, each metric evaluator ran three times, and we considered the median value as the final metric value.

\promesse~\cite{Mapomme15b} takes an $\alpha$ parameter (expressed in meters) specifying the distance to enforce between two consecutive locations (the larger $\alpha$, the higher the spatial distortion of the raw trace).
In \accio, $\alpha$ takes values in $[0, 500]$ (meters), compared to an $\alpha \in \{100,200,300,500\}$ for the static baselines.
Baselines allow use to explore different values regularly spaced, including $500$ meters, the maximum $\alpha$ considered by \accio and $200$ meters, that $\alpha$ that should be globally optimal, according to~\cite{Mapomme15b}.
Similarly to \geoi, we also set as objectives for the optimizer to minimise the POIs retrieval with the same setting, but we set to maximise the area coverage for the utility metric.
For the area coverage metric, we use Google's S2 geometry library\footnote{\url{https://github.com/google/s2-geometry-library-java}} to implement the $cell$ function and generate cells with a size covering a few blocks inside a city. 
This library is able to generate cells from a latitude and longitude at different levels, with the interesting property of cells having a similar area wherever they are on the globe.
We consider cells at the 15th level, areas at this level typically covering a few blocks inside a city.

\subsection{LPPM comparison}\label{subsec:offline}

We evaluate the offline scenario by using \accio to optimise both \geoi and \promesse in order to protect the Geolife dataset.
In this scenario, the system designer configures \accio to have a single value of $\epsilon$ and $\alpha$ (for \geoi and \promesse, respectively) for each user, and to evaluate these LPPMs through three metric evaluators: the POIs retrieval, the spatial distortion and the area coverage.
Nevertheless, our framework also allows the system designer to perform pre-processing on the dataset to split it into smaller data portions and to choose to tune the LPPM configuration for each data portion for instance.

Figure~\ref{fig:offline_eval} reports the Cumulative Distribution Function (CDF) of the POIs retrieval, the spatial distortion and the area coverage for both LPPMs.
For all these metric evaluators, the configuration found by \accio for \promesse provides better results than the configuration found by \accio for \geoi.
Indeed, more than 95~\% of users using \promesse have a POI retrieval of 0 (i.e., all of their POIs are hidden), a median spatial distortion of 25 meters (respectively 75 meters for \geoi) and a median area coverage of 0.75 (respectively 0.55 for \geoi).
Ultimately, the decision is left to the system designer to select which of the resulting protected dataset she would use. 
\accio only provides all the necessary material to easily evaluate LPPMs according to privacy and utility objectives.


\begin{figure*}[tb]
  \centering
  \subfloat[\geoi -- Geolife dataset]{\label{fig:geoind_epsilon_cdf-geo}\includegraphics[width=0.35\textwidth]{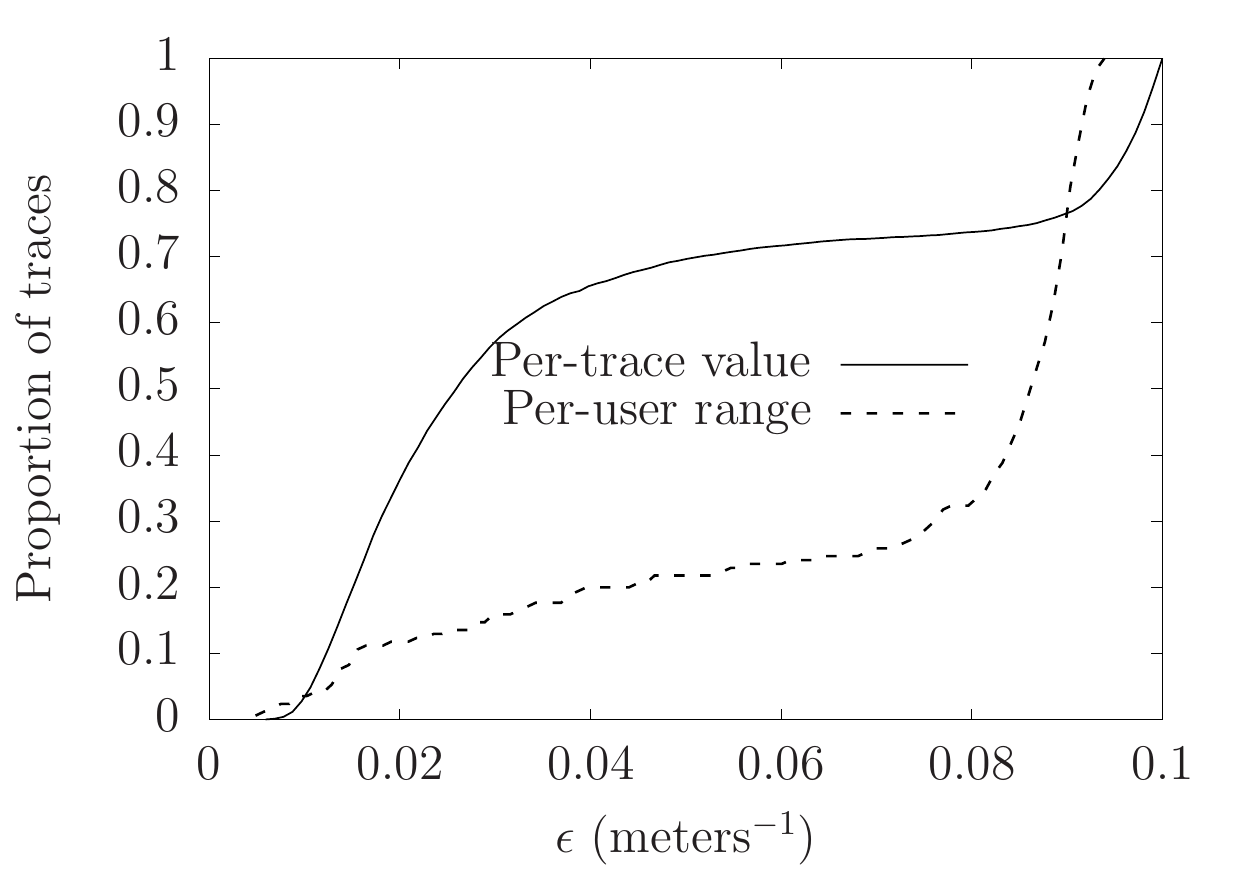}}
  \subfloat[\geoi -- MDC dataset]{\label{fig:geoind_epsilon-cdf-mdc}\includegraphics[width=0.35\textwidth]{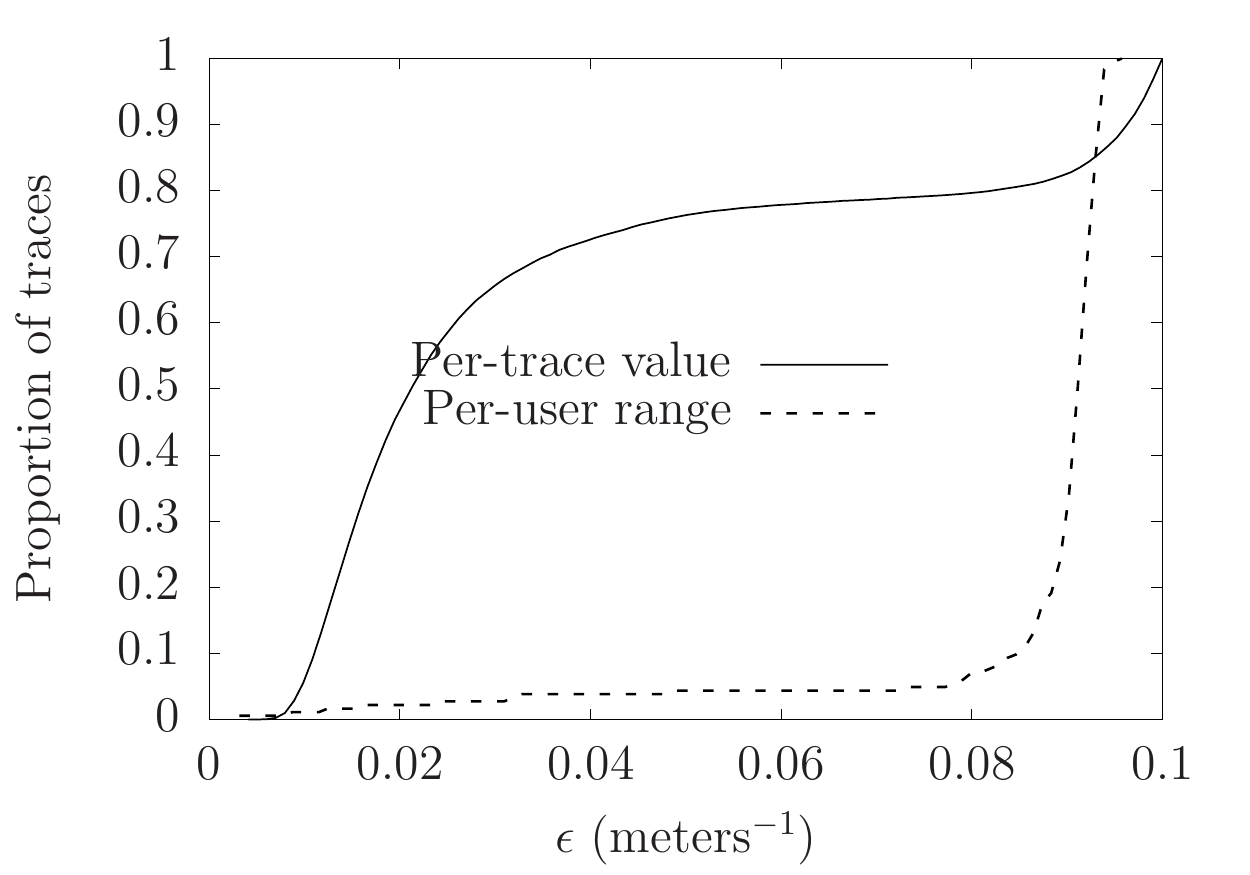}}

  \subfloat[\promesse -- Geolife dataset]{\label{smooth_epsilon_cdf-geo}\includegraphics[width=0.35\textwidth]{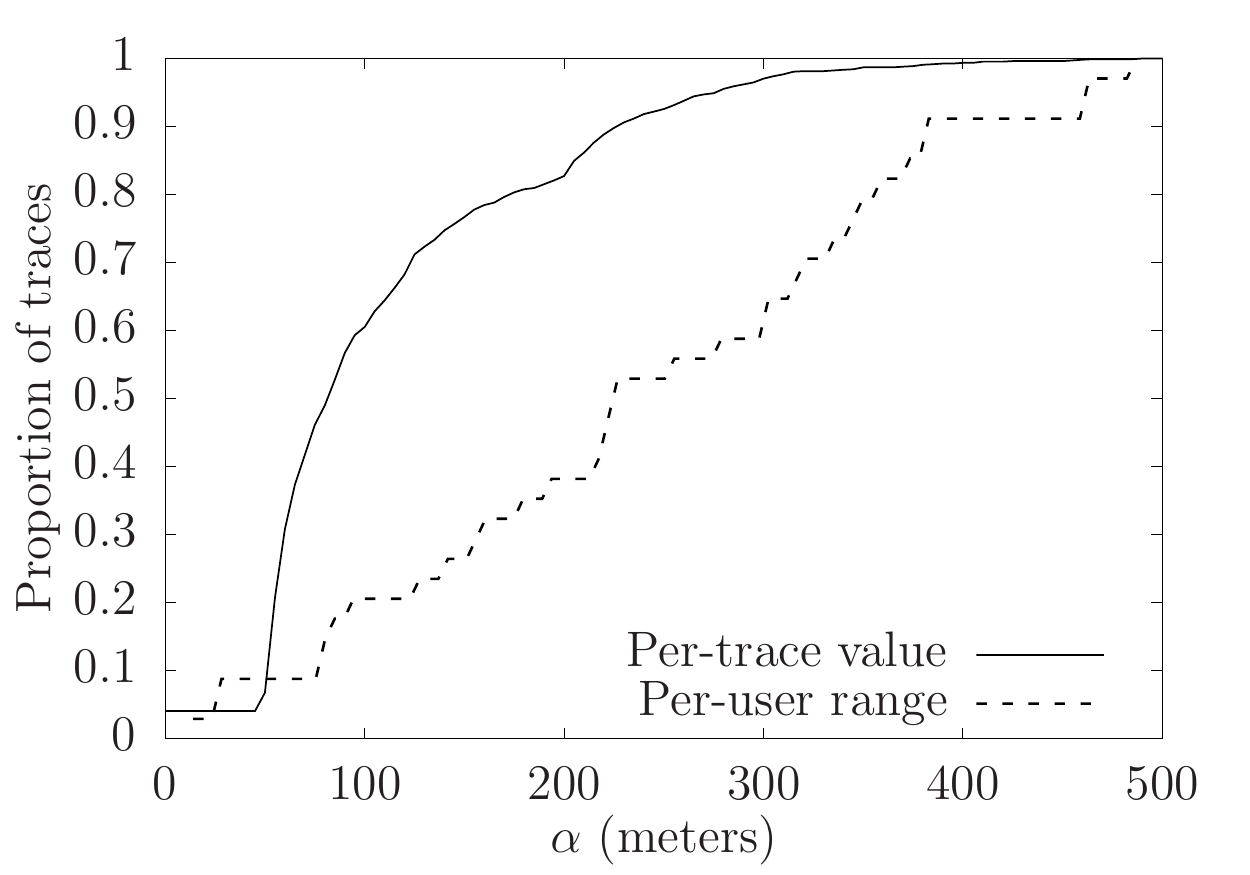}}
  \subfloat[\promesse -- MDC dataset]{\label{smooth_epsilon_cdf-mdc}\includegraphics[width=0.35\textwidth]{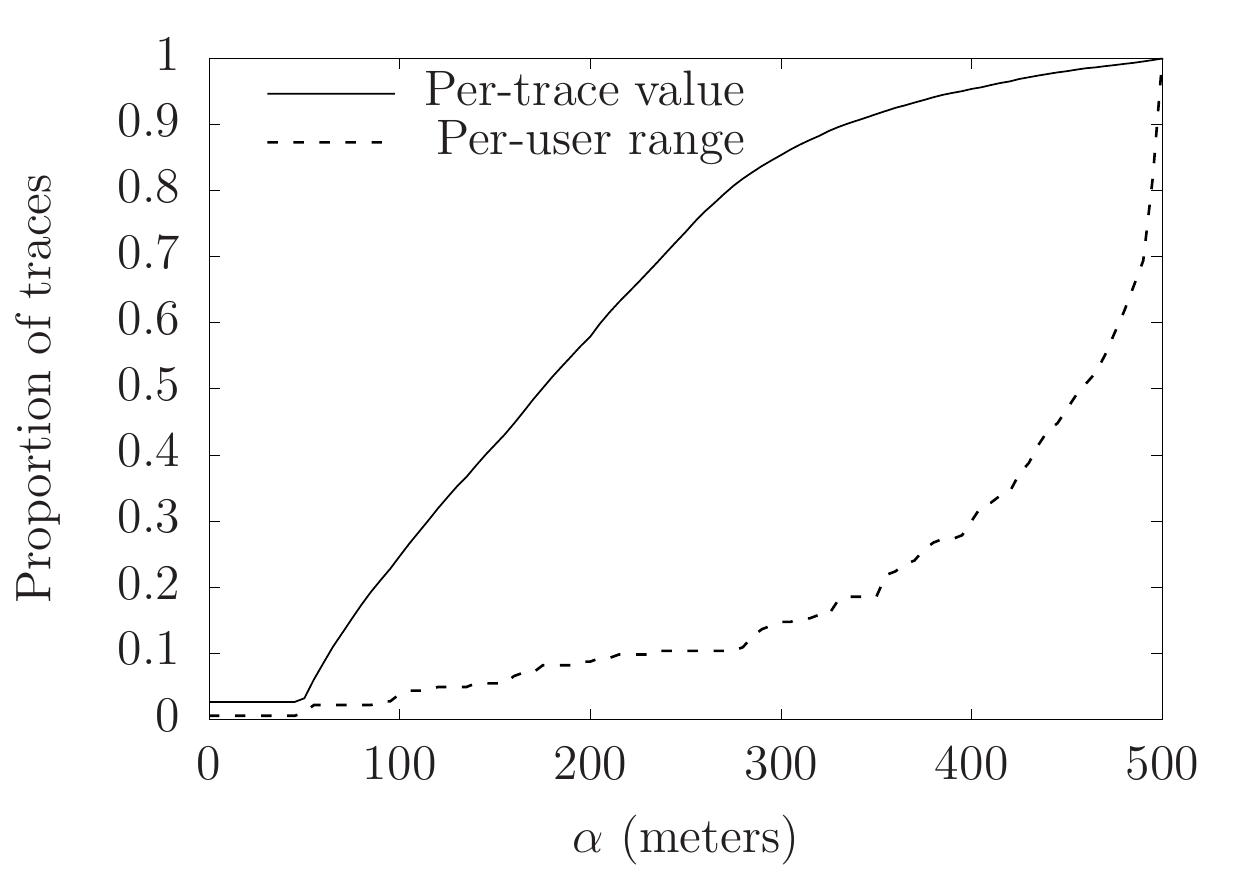}}

  \caption{Cumulative distribution function of the value taken by $\epsilon$ and $\alpha$ for \geoi and \promesse, respectively.}
  \label{fig:epsilon_cdf}
\end{figure*}

\subsection{Privacy and utility trade-off}\label{subsec:tradeoff}

We now illustrate the online scenario.
We consider a crowd sensing application that collects the user location every 30 seconds through his mobile device, and sends this data once a day to an LBS.

Figure~\ref{fig:geoind_eval} reports for \geoi and for the two considered datasets the CDF of the privacy and the utility objective metrics (i.e., POIs retrieval and spatial distortion, respectively) for both the dynamic configuration of $\epsilon$ found by \accio and several static configurations of $\epsilon$.
We show that \accio hides all POIs of at least 92~\% of users (i.e., a null F-Score) for both datasets (Figures~\ref{fig:geoind_eval-pgeo}-\ref{fig:geoind_eval-pmdc}) 
while maintaining a median spatial distortion of 40 and 70 meters with Geolife and MDC, respectively (Figures~\ref{fig:geoind_eval-pgeo}-\ref{fig:geoind_eval-pmdc}).
Note that some static $\epsilon$ configurations outperform our dynamic solution either on privacy or on utility (e.g., the one with the lowest value of $\epsilon$ is better for hiding POIs but has a worse spatial distortion and the one with the highest value of $\epsilon$ has the opposite behaviour).
Nevertheless, there is no static configuration that outperforms the dynamic configurations found by \accio both on privacy and on utility.
This means that the trade-off between privacy and utility provided by \accio is better than one found by the static baselines as the latter adjusts the amount of noise according to the underlying data to protect.

Figure~\ref{fig:promesse_eval}, in turn, depicts for \promesse using the Geolife and MDC datasets, the CDF of the privacy and the utility 
objective metrics (i.e., POIs retrieval and area coverage,  
respectively) for both the dynamic configuration of \accio and static baselines.
We show that the dynamic $\alpha$ configuration of \accio offers a nearly perfect protection with a null F-Score for almost all users and on both datasets (Figures~\ref{fig:pro_eval-pgeo}-\ref{fig:pro_eval-pmdc}), while offering a better utility (i.e., the smaller area coverage) than the various static $\alpha$ configurations (Figures~\ref{fig:pro_eval-pgeo}-\ref{fig:pro_eval-pmdc}).

In the case of \promesse, these results show that \accio is able to provide the best of the two worlds by outperforming static configurations both on privacy and utility.

\subsection{Adaptive configuration}\label{subsec:adaptive}

We now focus our evaluation on the analysis of the adaptive capabilities of \accio.
Specifically, we analyse the variation of the LPPM parametrisation according to the evolution of the input trace under analysis.
Figure~\ref{fig:epsilon_cdf} shows for both \geoi and \promesse on the two considered datasets the CDFs of the different values of $\epsilon$ or $\alpha$ generated by \accio for each trace, and the range of parameter values taken for each user (i.e., max - min).

Interesting enough, results for \geoi (Figures \ref{fig:geoind_epsilon_cdf-geo}-\ref{fig:geoind_epsilon-cdf-mdc}) show that 65~\% with Geolife (respectively 72~\% with MDC) of the chosen per-trace values for $\epsilon$ are smaller than 0.04, and 27~\% (respectively 20~\%) are greater than 0.09.
Values of $\epsilon$ between 0.04 and 0.08 are rarely chosen by our algorithm, which could indicate that either a trace needs to be strongly protected or almost not. 
If we consider the range of $\epsilon$ values taken per-user, results show that for 77~\% of users with Geolife (respectively 93~\% with MDC) the range of values is greater than 0.08 (out of a maximum of 0.1).
This large range indicates that \accio chooses very different values of $\epsilon$ for each user during their mobility activity.
This variability across traces of a single user highlights the dynamic optimisation that \accio performs to adapt the configuration parameter of the protection mechanism according to the data portion under analysis.

Results for \promesse (Figures \ref{smooth_epsilon_cdf-geo}-\ref{smooth_epsilon_cdf-mdc}) exhibit a different behaviour.
The different values chosen by $\alpha$ per-trace are almost chosen uniformly distributed across the range of possible values, with a median value of 80 and 170 meters with Geolife and MDC, respectively.
The range of $\alpha$ values taken for each user also reports a uniform distribution for Geolife. 
However, the per-user range for MDC exhibits a different distribution where 70~\% of users have a range greater than 400 meters (out of 500 meters).
For both datasets, the large range chosen for $\alpha$ supports once again the necessity to adapt configuration parameters of LPPMs according to the current mobility data.

\subsection{Deployment on mobile devices}\label{subsec:mobile}

Finally, we evaluate the cost of running \accio on a mobile device. 
More precisely, we measure the time taken by a mobile device to perform the optimisation of the configuration parameters of an LPPM. 
Depending on the considered use case, for non real-time scenarios (e.g., periodically sending batches of data), the impact of the introduced latency remains low.
However, this latency must be limited to avoid the user device to be frozen while the optimizer is running. 
To achieve this measurement, we constrained this particular experiment to run on a single core, clocked to 1.2 GHz, and with 1 Go of RAM.
The time taken by \accio to find a parametrisation in this case is on average of 9 seconds with \geoi and 500ms with \promesse.
We found that the rate at which we collect records has a non-negligible impact on the performance.
For instance, if we collect a record every 5 minutes (instead of 30 seconds in the current experiments), the execution time with \geoi is on average of 7 seconds (22~\% less) due to a smaller size of the batch of data to be processed.

\section{Conclusion}
\label{sec:conclusion}
In this paper, we presented \accio, an adaptive location privacy framework.
\accio makes the parametrisation and the evaluation of location privacy protection mechanisms easier by shifting the process of protecting location information from a parameter-centric paradigm where users or system designers have to set obscure parameters, to an objective-centric paradigm where users only have to define their target privacy and utility objectives.
Using these objectives, \accio automatically tunes the set of LPPM configuration parameters according to the data under analysis, which allows adding the right amount of noise and avoids unnecessarily degrading the quality of the data or under protecting sensitive data portions. 
We illustrated the capabilities of our framework through the optimisation of two state-of-the-art LPPMs on two use case scenarios and by relying on two real datasets. 
We showed that \accio enabled finding dynamic LPPM configurations that outperform representative static ones thus reaching the best of both worlds in terms of privacy and utility.
Extending our framework to support more protection mechanisms, metric evaluators, and objectives is part of our future work.
In addition, designing and integrating to \accio collaborative protection mechanisms (e.g., \cite{Ghinita07}) as well as privacy attacks leveraging the full knowledge of the mobility traces (e.g., the user re-identification attack defined in~\cite{Mapomme14}) is also an interesting research perspective.



\section*{Acknowledgment}
This work was supported by the LABEX IMU (ANR-10-LABX-0088) of Université de Lyon, within the program "Investissements d'Avenir" (ANR-11-IDEX-0007) operated by the French National Research Agency (ANR).
Portions of the research in this paper used the MDC Database made available by Idiap Research Institute, Switzerland and owned by Nokia.

\bibliographystyle{IEEEtran}
\bibliography{IEEEabrv,bibli}

\begin{thebibliography}{10}
\providecommand{\url}[1]{#1}
\csname url@samestyle\endcsname
\providecommand{\newblock}{\relax}
\providecommand{\bibinfo}[2]{#2}
\providecommand{\BIBentrySTDinterwordspacing}{\spaceskip=0pt\relax}
\providecommand{\BIBentryALTinterwordstretchfactor}{4}
\providecommand{\BIBentryALTinterwordspacing}{\spaceskip=\fontdimen2\font plus
\BIBentryALTinterwordstretchfactor\fontdimen3\font minus
  \fontdimen4\font\relax}
\providecommand{\BIBforeignlanguage}[2]{{%
\expandafter\ifx\csname l@#1\endcsname\relax
\typeout{** WARNING: IEEEtran.bst: No hyphenation pattern has been}%
\typeout{** loaded for the language `#1'. Using the pattern for}%
\typeout{** the default language instead.}%
\else
\language=\csname l@#1\endcsname
\fi
#2}}
\providecommand{\BIBdecl}{\relax}
\BIBdecl

\bibitem{Henttu12}
H.~Henttu, J.-M. Izaret, and D.~Potere, ``{Geospatial Services: A \$1.6
  Trillion Growth Engine for the U.S. Economy},''
  \url{http://www.bcg.com/documents/file109372.pdf}, 2012.

\bibitem{Gambs11}
S.~Gambs, M.-O. Killijian, and M.~N. del Prado~Cortez, ``{Show Me How You Move
  and I Will Tell You Who You Are},'' \emph{Transactions on Data Privacy},
  vol.~4, no.~2, pp. 103--126, Aug. 2011.

\bibitem{Sharad14}
K.~Sharad and G.~Danezis, ``An automated social graph de-anonymization
  technique,'' in \emph{WPES}, 2014, pp. 47--58.

\bibitem{Sadilek12}
A.~Sadilek and J.~Krumm, ``Far out: Predicting long-term human mobility,'' in
  \emph{AAAI on Artificial Intelligence}, 2012.

\bibitem{Franceschi15}
L.~Franceschi-Bicchierai, ``Redditor cracks anonymous data trove to pinpoint
  muslim cab drivers,''
  \url{http://mashable.com/2015/01/28/redditor-muslim-cab-drivers/}, Jan. 2015.

\bibitem{Acs14}
G.~Acs and C.~Castelluccia, ``{A Case Study: Privacy Preserving Release of
  Spatio-temporal Density in Paris},'' in \emph{SIGKDD}, 2014, pp. 1679--1688.

\bibitem{Andres13}
M.~E. Andr{\'e}s, N.~E. Bordenabe, K.~Chatzikokolakis, and C.~Palamidessi,
  ``{Geo-indistinguishability: Differential Privacy for Location-based
  Systems},'' in \emph{SIGSAC}, 2013, pp. 901--914.

\bibitem{Mapomme15b}
V.~Primault, S.~Ben~Mokhtar, C.~Lauradoux, and L.~Brunie, ``{Time Distortion
  Anonymization for the Publication of Mobility Data with High Utility},'' in
  \emph{TrustCom}, 2015.

\bibitem{Fawaz14}
K.~Fawaz and K.~G. Shin, ``Location privacy protection for smartphone users,''
  in \emph{SIGSAC}, 2014.

\bibitem{Shokri11}
R.~Shokri, G.~Theodorakopoulos, J.-Y. Le~Boudec, and J.-P. Hubaux,
  ``Quantifying location privacy,'' in \emph{SSS}, 2011, pp. 247--262.

\bibitem{Agir14}
B.~Agir, T.~Papaioannou, R.~Narendula, K.~Aberer, and J.-P. Hubaux,
  ``{User-side adaptive protection of location privacy in participatory
  sensing},'' \emph{GeoInformatica}, vol.~18, no.~1, pp. 165--191, 2014.

\bibitem{Chatzikokolakis15}
K.~Chatzikokolakis, C.~Palamidessi, and M.~Stronati, ``{Constructing elastic
  distinguishability metrics for location privacy},'' in \emph{PETS}, vol.
  2015, 2015, pp. 156--170.

\bibitem{privamov-srds16}
``{ALP: Adaptive Location Privacy},''
  \url{http://liris.cnrs.fr/privamov/project/publications/srds16}.

\bibitem{Sweeney02}
L.~Sweeney, ``{k-Anonymity: A model for protecting privacy},''
  \emph{{International Journal of Uncertainty, Fuzziness and Knowledge-Based
  Systems}}, vol.~10, no.~5, pp. 557--570, 2002.

\bibitem{Dwork06}
C.~Dwork, ``{Differential Privacy},'' in \emph{{Automata, Languages and
  Programming}}, ser. Lecture Notes in Computer Science.\hskip 1em plus 0.5em
  minus 0.4em\relax Springer Berlin Heidelberg, 2006, vol. 4052, pp. 1--12.

\bibitem{Mokbel06}
M.~F. Mokbel, C.-Y. Chow, and W.~G. Aref, ``{The New Casper: Query Processing
  for Location Services Without Compromising Privacy},'' in \emph{VLDB}, 2006,
  pp. 763--774.

\bibitem{Ghinita07}
G.~Ghinita, P.~Kalnis, and S.~Skiadopoulos, ``{PRIVE: Anonymous Location-based
  Queries in Distributed Mobile Systems},'' in \emph{WWW}, 2007, pp. 371--380.

\bibitem{Krumm07}
J.~Krumm, ``{Inference Attacks on Location Tracks},'' in \emph{Pervasive},
  2007, pp. 127--143.

\bibitem{Kirkpatrick82}
S.~Kirkpatrick, C.~Gelatt, and M.~P. Vecchi, ``Optimization by simulated
  annealing,'' \emph{Science}, vol. 220, no. 4598, pp. 671--680, May 1983.

\bibitem{Mapomme14}
V.~Primault, S.~Ben~Mokhtar, C.~Lauradoux, and L.~Brunie, ``{Differentially
  Private Location Privacy in Practice},'' in \emph{MoST}, 2014.

\bibitem{Hariharan04}
R.~Hariharan and K.~Toyama, ``{Project Lachesis: parsing and modeling location
  histories},'' in \emph{{Geographic Information Science}}, 2004, pp. 106--124.

\bibitem{Abul08}
O.~Abul, F.~Bonchi, and M.~Nanni, ``{Never Walk Alone: Uncertainty for
  Anonymity in Moving Objects Databases},'' in \emph{{ICDE}}, 2008, pp.
  376--385.

\bibitem{Zheng10}
Y.~Zheng, X.~Xie, and W.-Y. Ma, ``{GeoLife: A Collaborative Social Networking
  Service among User, location and trajectory},'' \emph{Data Engineering
  Bulletin}, vol.~33, no.~2, pp. 32--40, 2010.

\bibitem{Kiukkonen10}
N.~Kiukkonen, J.~Blom, O.~Dousse, D.~Gatica-perez, and J.~Laurila, ``{Towards
  rich mobile phone datasets: Lausanne data collection campaign},'' in
  \emph{ACE}, 2010.

\end{thebibliography}

\end{document}